# ChatGPT at the Speed of Light: Optical Comb-Based Monolithic Photonic-Electronic Linear-Algebra Accelerators

Tzu-Chien Hsueh, *Senior Member, IEEE*, Yeshaiahu Fainman, *Fellow, IEEE*, and Bill Lin, *Senior Member, IEEE*

***Abstract*—This paper proposes to adopt advanced monolithic silicon-photonics integrated-circuits manufacturing capabilities to achieve a system-on-chip photonic-electronic linear-algebra accelerator with the features of optical comb-based broadband incoherent photo-detections and high-dimensional operations of consecutive matrix-matrix multiplications to enable substantial leaps in computation density and energy efficiency, with practical considerations of power/area overhead due to photonic-electronic on-chip conversions, integrations, and calibrations through holistic co-design approaches to support attention-head mechanism based deep-learning neural networks used in Large Language Models and other emergent applications.**

***Index Terms*—Attention-head mechanism, ChatGPT, frequency comb, linear algebra, matrix-matrix multiplication, matrix-vector multiplication, micro resonator, monolithic integration, silicon-photonics, Transformer model.**

## I. Introduction

RECENT advances in ChatGPT [1], and its underlying deep machine learning model called the Transformer [2], have shown unprecedented capabilities in artificial intelligence (AI). ChatGPT (or simply GPT), an important and impressive Large Language Model (LLM) developed by OpenAI [1], has the ability to understand and generate text that appears to resemble human conversations. According to a recent study [3], ChatGPT has already become the fastest-growing consumer application in history, having reached over 100 million active monthly users in just two months after launch. Beyond the most obvious impact of ChatGPT in its potential to transform communication and information, it is also capable of writing code in popular languages such as Python, C, Java, and JavaScript [4]. Also, the latest version of GPT was able to score near the 90-th percentile on an actual bar exam [5]. Beyond language-based applications, attention-head mechanism [2] has been applied successfully to scientific problems that on the surface seem quite far from the initial intent of Transformer models developed for natural language processing. One notable example application is the protein-folding problem that is well-known to require astronomical amounts of computing power using traditional computational chemistry approaches. In particular, protein-folding aims to predict precise three-dimensional structures of proteins, which is essential for understanding their function, designing new drugs, and advancing our understanding of diseases and biological systems. In a groundbreaking paper [6], the authors demonstrated AlphaFold that is able to predict three-dimensional protein structures with remarkable accuracy as a generative AI task, based on the same underlying generative model that powers GPT.

To realize Transformer models in the hardware physical layers with such high computation capacities as mentioned, photonic computing via monolithic silicon-photonics (SiPh) fabrication and integration [7], [8] is the primary implementation strategy of the proposed linear-algebra accelerator based on the major evidence as follows. First, since modern data bandwidth requirements and the standardization of SiPh integrated circuits (IC), photonic technology has been widely used for both long- and short-reach high-volume data communications [9], [10], [11], [12], [13], [14]. Meanwhile, due to the advancement of deep learning models far outpacing the Moore's law [15] and energy/area bottleneck of classical von Neumann computing architectures, photonic devices and ICs, possessing inherent parallelism, high degree of connectivity, and speed-of-light propagation associated with wavelength-division-multiplexing (WDM) technique [27] in the optical communication, have been broadly adopted in the computation tasks of linear operations such as passive Fourier transforms [16] and matrix operations [15], [17], [18], [19], [20], [21], [22], [23], [24] exhibiting superior advantages of photonic computing in bandwidth density, processing latency, silicon area, and energy consumption. Especially, the concept of optical comb generations [40] plays the key role in the broadband incoherent photo-detection to cover more than 128 carriers with 30 GHz to 80 GHz frequency spacing across the The authors are with the Department of Electrical and Computer Engineering, University of California, San Diego, La Jolla, CA 92093 USA (e-mail: tzhsueh@ucsd.edu).

entire WDM spectrum for the proposed linear-algebra accelerator. Second, the availability of commercial monolithic SiPh semiconductor process technology, e.g., GlobalFoundaries 45SPCLO [7], [8], represents an exciting opportunity to explore holistic co-design approaches that leverage unique capabilities of photonics and CMOS electronics to advance the state of computing that is currently at a crossroad. Particularly, monolithic SiPh technology consolidating all required photonic and electronic devices/circuits for the proposed linear-algebra accelerator on a single die can tremendously eliminate power/area/integration overhead due to interfacing I/O circuits (SerDes and data transceivers), electrostatic-discharge (ESD) protection diodes, chip bumps/pads, and interposers/packages among separate photonic and electronic dies, which mostly have been ignored in the performance metrics of photonic computing literatures [15], [18], [19], [20], [21], [22], [23], [24] but exposed by the limited computation scales and dimensions of their realistic hardware demonstrations.

The goal of the proposed monolithic photonic-electronic (MPE) linear-algebra accelerators is to exploit a new application of optical combs and practically realize a well-interfaced and power/area efficient system-on-chip (SoC) possessing the functionality of high-dimensional matrix-vector multiplications (MVM), matrix-matrix multiplications (MMM), and double matrix-matrix multiplications (D-MMM) for the attention-head mechanism in Transformer models. The remainder of the paper is organized as follows. The background of attention-head mechanism in Transformer models is summarized in Section II. The motivation of using monolithic SiPh technology and the attention-head optimization for the MPE realization are elaborated in Section III. The architecture and building-block functionality with performance specifications of an MPE-MVM accelerator are analyzed in Section IV. The architecture scalability and parallelism of MPE-MMM and MPE-D-MMM accelerators are described in Section V and VI, respectively. The performance evaluation and conclusion are summarized in Section VII.

## II. Background of Transformers

The Transformer model [2] is a type of deep learning network architecture that employs self-attention as a mechanism for processing input sequences. It can efficiently capture long-range dependencies and relationships within the data by differentially weighting the significance of each part of the input sequence. The Transformer model achieves this by means of a building block called a *scaled dot-product attention unit* that calculates attention weights simultaneously between all input tokens. This calculation produces embeddings for every token in context that contain information about the token itself, along with a weighted combination of other relevant tokens based on their respective attention weights. In particular, the Transformer model learns three weight matrices for each attention unit: the query weights $W_Q$, the key weights $W_K$, and the value weights $W_V$. The Query, Key, and Value matrices can then be generated by multiplying the input sequence $X = [x_0, x_1, \ldots, x_n]$ with the corresponding trained weight matrices:

$$Q = X \cdot W_Q; \quad K = X \cdot W_K; \quad V = X \cdot W_V \tag{1}$$

Given these matrices, the attention score can be calculated using the following scaled dot-product attention equation:

$$Attention(Q, K, V) = softmax\left(\frac{Q \cdot K^T}{\sqrt{d_k}}\right) \cdot V \tag{2}$$

where $d_k$ represents the dimension of the keys and values. As shown in Eq. (1) and (2), the central computations involve high-dimensional MVMs and MMMs.

The calculations in Eq. (1) and (2) basically form a single *attention-head*, with the corresponding set of weight matrices $(W_Q, W_K, W_V)$. In the Transformer model proposed in [2], each layer has *multiple* attention-heads. With multiple attention-heads, each head can *attend* to a different notion of *relevance* by learning a different set of projection matrices $(W_Q^i, W_K^i, W_V^i)$ for each attention-head $i$. The computations for each attention-head can then be executed in parallel to enable rapid processing. The outputs of $h$ attention-heads are then concatenated and passed into a feedforward network layer as follows:

$$\begin{aligned} &Multihead(Q, K, V) \\ &= Concat\left[Attention\left(X \cdot W_Q^i, X \cdot W_K^i, X \cdot W_V^i\right)\right] \cdot W_O \end{aligned} \tag{3}$$

where $W_O$ is the final projection matrix for the entire multi-headed attention unit. The overall Transformer architecture makes use of these multi-headed attention units as basic building blocks for encoders and decoders. Encoders and decoders can then be stacked together to provide increasing learning capacities to capture long-range dependencies and relationships within the input sequence.

## III. Optimization between Attention-Heads & Monolithic Photonic-Electronic Accelerators

Though a number of developments have explored photonic accelerators for convolutional neural networks (CNNs) and recurrent neural networks (RNNs) [18], [19], [20], [21], [22], [23], [24] with promising results, many substantial challenges must be overcome when considering photonic computing for the attention-head calculations in Transformer workloads with such high-dimensional MMM computations. First, prior work on photonic accelerators for CNNs and RNNs have focused on MVM computations that are typically performed with *static* weights that do not require reprogramming once learned. However, the compute-kernels in a Transformer network are different: Transformer models require MMMs between the Query, Key, and Value matrices that are *dynamically* generated from *changing* inputs. Second, the attention-head calculations involve not only high-dimensional but also consecutive MMMs. If implemented in a naïve way with MVMs, the linear projection of inputs and the calculation of the scaled dot-product attention could require significant number of conversions between the photonic and electronic domains. Third, prior work on photonic accelerators have largely assumed separate chip implementations for the photonic and electronic parts – these approaches incur very expensive chip-to-chip communications and integrations that decrease the effectiveness of a photonic computing approach to deep learning acceleration.

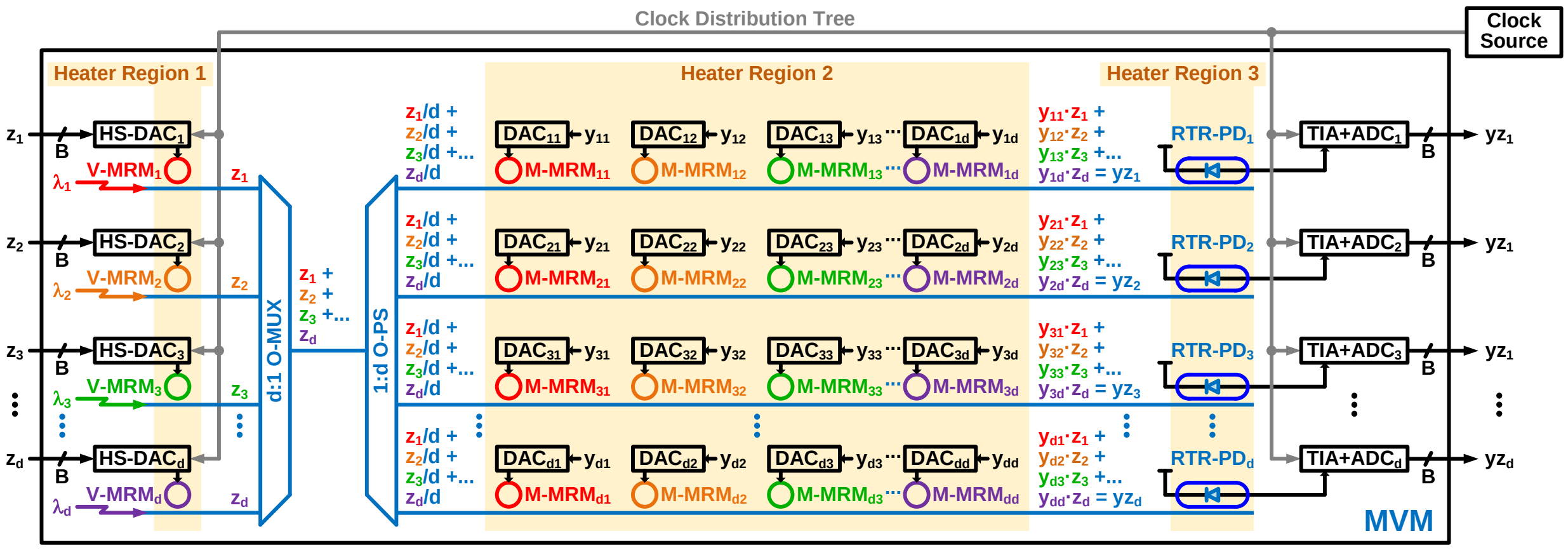

Fig. 1. The system block diagram of the MPE-MVM linear-algebra accelerator. Note that each "yz" presents a "single-word" variable, NOT "y times z".

The innovations and approaches in Section IV, V, and VI aim to address these substantial challenges by performing MVMs, MMMs and D-MMMs purely in the photonic domain of a single monolithic SiPh chip without "intermediate" optical memory and O/E/O conversions to demonstrate the feasibility and capabilities of the MPE linear-algebra accelerator for executing Transformer-model workloads with two layers of power/area/speed enhancements: first, the computation end-to-end E/O and O/E cost reduction through the monolithic SiPh integration; second, the number of O/E/O reduction through the attention-head architecture innovation, which is algorithmically elaborated in the rest of this section.

As shown in Eq. (1) and (2), the attention-heads involve a significant number of MMM computations, including the linear projections of the input vector by three trained matrices to produce Query, Key, and Value matrices, followed by a MMM of the Query and Key matrices and another MMM with the Value matrix. This paper proposed to co-design the attention-head computations in a holistic manner together with the capability of the MPE linear-algebra accelerator as mentioned. The first idea is to collapse the linear projections of the Query and Key matrices and their multiplication. In particular, $(Q \cdot K^T)$ can be rewritten as follows:

$$\begin{aligned} C = Q \cdot K^T &= \left(X \cdot W_Q\right) \cdot \left(X \cdot W_K\right)^T \\ &= \left[X \cdot \left(W_Q \cdot W_K^T\right)\right] \cdot X^T = \left(X \cdot W_C\right) \cdot X^T \end{aligned} \tag{4}$$

where $W_C = W_Q \cdot W_K^T$ can be computed offline since the entries in both $W_Q$ and $W_K$ are constants once trained. As shown in Eq. (4), collapsing the linear projections results in a *double matrix-matrix multiplication* (D-MMM), which can be realized by an MPE-D-MMM accelerator architecture described in Section VI, *without intermediate O/E/O conversions* between two individual MMMs. Once matrix $C$ in Eq. (4) is computed as a complete D-MMM result, the outcome is converted from photonic to electronic domain, and then the scaling by $1/\sqrt{d_k}$ and $softmax(\cdot)$ required in each attention-head can be efficiently realized as simple digital shifts and table lookup operations in the electronic domain. In particular, $\sqrt{d_k}$ is typically chosen to be some $2^m$ so scaling $c_{ij} \in C$ by $1/\sqrt{d_k}$ can be performed simply by a right-shift of $c_{ij}$ by $m$ bits. As shown in [25] and [26], the $softmax(\cdot)$ function can be rewritten as follows:

$$\begin{aligned} softmax(a_i) &= \frac{e^{a_i}}{\sum_{j=1}^{K} e^{a_j}} \\ &= \exp\left[a_i - a_{max} - \log\left(\sum_{j=1}^{K} e^{a_j - a_{max}}\right)\right] \end{aligned} \tag{5}$$

where $a_{max}$ is the maximum among the $a_i$ values. Also, as discussed in [25] and [26], $\exp(\cdot)$ and $\log(\cdot)$ can be efficiently approximated with table lookups, where the quantization of the $a_i$ values is co-designed with the MPE accelerator to optimize for the best tradeoffs. With matrix $S = softmax\left(C/\sqrt{d_k}\right)$ computed as right-shifts and table lookups in accordance with Eq. (5), the remaining attention-head computation is another D-MMM of three matrices:

$$Attention = S \cdot V = S \cdot (X \cdot W_V) \tag{6}$$

which can again take advantage of the MPE-D-MMM accelerator to avoid the intermediate O/E/O conversion within the operation of Eq. (6).

## IV. MPE Matrix-Vector Multiplications

A unified MPE-MVM accelerator serves as the key building block of the MPE-MMM, MPE-D-MMM, and eventually attention-heads in the Transformer model with high degree of reconfigurability in terms of the internal matrix weights and accelerator-to-accelerator on-chip physical connections. Meanwhile, the primary MVM functionality is realized by exploiting high degree of freedom in spatial parallelism with the WDM technique [27]. As shown in Fig. 1, each high-dimensional MPE-MVM accelerator consists of "d" B-bit high-speed digital-to-analog converters (HS-DAC), "$d^2$" low-power static DACs, "d" transimpedance amplifiers (TIA) individually followed by "d" B-bit analog-to-digital converters (ADC), digital registers/logics, clock distribution, and discrete-time iteration mechanism in the electronic domain as well as "d" vector micro-ring modulators (V-MRM) for the input vector E/O conversion, "$d^2$" matrix micro-ring modulators (M-MRM) for the matrix E/O conversion with photonic MVM operations, "d" optical comb-based racetrack-resonator photodetectors (RTR-PD), optical multiplexer (O-MUX), optical power splitter (O-PS), and waveguides in the photonic domain.

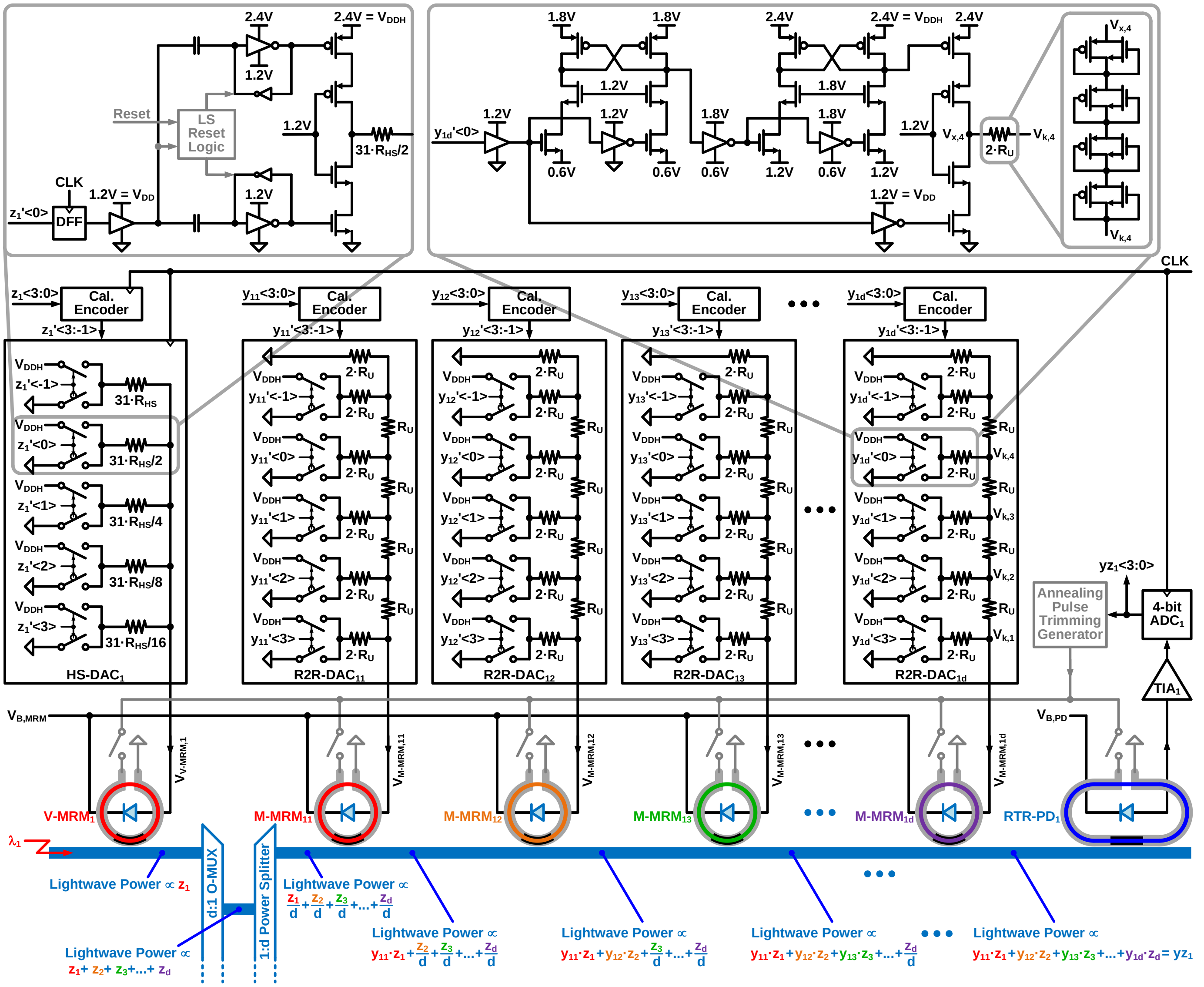

Fig. 2. The zoom-in version of the first MPE-MVM row from the input $z_1$<3:0> to output $yz_1$<3:0> with detailed circuits, interconnections, and clocking relation among the HS-DACs, R2R-DACs, MRMs, RTR-PDs, and post-fabrication trimming mechanism.

The input d-by-1 data vector is denoted by $Z_{d\times 1}$ with elements $z_i$, i = 1 ~ d; the output d-by-1 data vector is denoted by $YZ_{d\times 1}$ with elements $yz_i$, i = 1 ~ d (note that each "yz" presents a "single-word" variable, NOT "y times z"); the primary d-by-d matrix is denoted by $Y_{d\times d}$ with elements $y_{ij}$, both i and j = 1 ~ d individually. The MVM functionality is represented by $Y_{d\times d}\cdot Z_{d\times 1} = YZ_{d\times 1}$. The timings of both input and output vectors are managed by the electronic circuits in the digital domain to enable the flexibility of cascading and parallelizing the MPE-MVM accelerators and therefore to perform multi-stage MVMs while the MPE-MVM accelerator can also seamlessly interface with other on-chip digital circuits, processors, lookup tables, registers, and memory. The detailed circuits and interconnection within the MPE-MVM accelerator is shown in Fig. 2 as a zoom-in version of the first MVM row from the input $z_1$ to output $yz_1$. It is important to note that the clock-rate per MVM operation, playing one of the key roles for the computation throughput (MAC/s), is determined by the circuits/devices latency from the clock edge launching the HS-DAC through the MRM, O-MUX, O-PS, WDM-based MVM operation, RTR-PD, TIA, all the way to the ADC output data sampled by the next clock edge, which sets the clock period per MVM computation. Thus, the speeds of all electronic and photonic circuits/devices within this critical path determine the clock period (i.e., register-to-register clock cycle) and eventually computation throughput (MAC/s) of this MPE-MVM accelerator, especially the electronic circuits dominating the entire power/area/speed performance metrics.

### A. WDM-Based MVM Computation Architecture

The MVM functionality is established on the concept of WDM incoherent data transmission in optical communication systems [9], [12], [27]. The data carries are "d" of laser light-waves with identical light-wave power but individual wavelengths $\lambda_i$, i = 1 ~ d, with proper spacing between adjacent wavelengths to assure incoherent detections. After the D/A and E/O conversions through the $\text{HS-DAC}_i$ and $\text{V-MRM}_i$, respectively, each $z_i$ of the input data vector $Z_{d\times1}$ is correspondingly presented by the light-wave power of $\lambda_i$. More precisely, the power of $\lambda_i$ light-wave is linearly proportional to $z_i$ with a consistent scalar across all i = 1 ~ d as shown in Fig. 1 and bottom-left of Fig. 2. Without considering any nonideality, the O-MUX with "d" fan-ins merges all power-modulated light-waves into a single waveguide, whose aggregate light-wave power is proportional to $\sum_{i=1}^{d} z_i = (z_1 + z_2 + z_3 + \ldots + z_d)$, and then the O-PS evenly splits this aggregate light-wave power into its "d" fan-outs so that the light-wave powers in the waveguides of all MVM rows shall be identical and proportional to $\sum_{j=1}^{d} z_j/d = (z_1 + z_2 + z_3 + \ldots + z_d)/d$. Note that the wavelength index for the "i-th MVM row" is "j" not "i".

After the O-PS, the light-wave power contributed by all $\lambda_j$, j = 1 ~ d, in the i-th MVM row will sequentially go through the resonance effects of "d" M-MRMs (i.e., $\text{M-MRM}_{ij}$, j = 1 ~ d). However, each $\text{M-MRM}_{ij}$ controlled by $y_{ij}$ can only affect or modulate the $\lambda_j$ light-wave power ($\propto z_j/d$) when the indexes j of $y_{ij}$ and $z_j$ are matched. E.g., as shown in the bottom of Fig. 2 for the first row of the MVM operation (i = 1), all $\lambda_j$ light-wave powers ($\propto \sum_{j=1}^{d} z_j/d$) together pass though the resonance effect of $\text{M-MRM}_{11}$ (j = 1), but only $\lambda_1$ light-wave power ($\propto z_1/d$) gets modulated to be proportional to $(y_{11}\cdot z_1)$; the rest of $\lambda_j$, j = 2 ~ d, light-wave powers can be all-pass filtered through $\text{M-MRM}_{11}$ without any power change. By resonating through all $\text{M-MRM}_{1j}$, j = 1 ~ d, all $\lambda_j$ light-wave powers in the first MVM row can be respectively modulated according to $y_{1j}$, j = 1 ~ d, at the speed of light. Right before reaching the $\text{RTR-PD}_1$, the total power on the first waveguide becomes proportional to $\sum_{j=1}^{d} y_{1j}\cdot z_j = (y_{11}\cdot z_1 + y_{12}\cdot z_2 + y_{13}\cdot z_3 + \ldots + y_{1d}\cdot z_d) = yz_1$, which is equivalent to the dot-product presentation of the input vector $Z_{d\times1}$ and the first-row vector of matrix $Y_{d\times d}$. By replicating the same process across all row vectors of $Y_{d\times d}$ in parallel, the total light-wave power of each waveguide can be eventually proportional to $yz_i$ with a consistent scalar across all i = 1 ~ d. After the following O/E and A/D conversions through $\text{RTR-PD}_i$ and $\text{ADC}_i$, respectively, the MVM operation is completed and all elements ($yz_i$, i = 1 ~ d) of output vector $YZ_{d\times1}$ is preserved in the electronic domain.

The similarity between WDM communication and WDM computation is that both rely on multiple wavelengths to independently carry their own signal/data information and to simultaneously propagate through a common communication channel to maximize the communication capacity or computation parallelism. On the other hand, there are two major differences. First, each light-wave power is only modulated once in WDM communication, which is merely for E/O conversion, but in WDM computation each light-wave power requires to be modulated at least "twice" to perform the E/O conversions and equivalent effect of "multiplications". Second, the receiver side of WDM communication needs to distinguish and separate all wavelengths and then detect each light-wave power individually to recover the data carried by each wavelength. Though WDM computation doesn't need to explicitly separate all wavelengths, the wavelength spectrum spacing still needs to be maintained because the "multiplication" is done when all light-wave powers are simultaneously present in the waveguide. Also, the equivalent summation of the dot-product requires consistent power-abortion photo-detections across the entire WDM bandwidth to maintain the computation integrity (or accuracy) during the O/E conversions. These two major differences between WDM communication vs. WDM computation dominantly determine the achievable E/O/E modulation/detection speeds (≥16 Gb/s vs. ≤4 Gb/s) and D/A/D data resolutions (≤2 bits vs. ≥4 bits) of these two optical WDM-based systems.

### B. High-Speed Digital-to-Analog Data Converters

In the primary computation path, each element $z_i$ of the input data vector $Z_{d\times1}$ represented by B-bit digital data (e.g., 4-bit in Fig. 2) is firstly converted into an analog-voltage level through an HS-DAC to drive a V-MRM for modulating the light-wave power of $\lambda_i$ as the E/O conversion process. To accommodate a reasonable E/O dynamic range (DR), the non-linear transmission power-gain induced by the MRM across all possible $2^B$ voltage-levels (e.g., $2^4 = 16$ levels in Fig. 2) can be alleviated by adding one calibration bit in the HS-DAC, so a basic digital calibration encoder is required for each HS-DAC mapping the incoming B-bit $z_i$ data to (B+1)-bit $z_i'$ data to perform better E/O conversion linearity while maintaining the same amount of $2^B$ voltage-levels (not $2^{B+1}$), which is further discussed in Section IV.D.

Each HS-DAC is implemented by a large-swing voltage-mode driver architecture [28] to maintain up to GHz sampling-rates (i.e., clock-rates), low static-power consumption, and high-voltage driving capability for maximizing the light-wave power DR per MRM. As shown in the left of Fig. 2, the binary segments controlled by $z_1'$<3:-1> in $\text{HS-DAC}_1$ have identical architecture, but the transistor-size (or driving capability) of each segment is reciprocally scaled according to its own series-resistor. In other words, all binary HS-DAC segments can be simply implemented by parallelizing multiple least-significant-bit (LSB) segments as shown in the top-left corner of Fig. 2 in a manner of powers of 2. E.g., the segment driven by $z_1'$<3> is formed by sixteen LSB segments driven by $z_1'$<-1>. Based on a certain binary combination of $z_1'$<3:-1>, some series-resistors can be shorted to $V_{DDH}$, and the rest to GND; thus, the HS-DAC output can generate a voltage-level based on the voltage-divider formed by all the parallel pull-up and pushed-down resistors between $V_{DDH}$ and GND. Note that two AC-coupled level-shifters are required in each HS-DAC for push-pull data latches [9] to enable high-speed, low-power, low-latency single-stage 2× voltage level-shifting from 1.2-V regular digital supply to 2.4-V high-voltage supply. Overall, to simultaneously maximize the E/O DR up to 2.4 V and maintain 45-nm CMOS reliability in 45SPCLO, separating power domains of the push-pull latches and adding cascade transistors in the push-pull driver paths are necessary to confine all CMOS devices operating within (1.2 to 2.4) V or (0 to 1.2) V.

The speed of the HS-DAC is determined by two factors. First, the latency the from clock edge of the $z_1$'<3:-1> register (i.e., DFF) to the push-pull switching transistors of the voltage-mode drivers, containing the delays of DFF clock-to-Q, digital data buffers, level-shifting AC-coupling capacitors, and push-pull data latches, is about 5× of a 20-ps fan-out of 4 (FO4) logic delay in 45SPCLO, i.e., 100 ps. Second, the RC time-constant of the HS-DAC output is determined by all parallel series-resistors from the standpoint of AC signal, i.e., a roughly constant and data-independent resistance ≈ ($31 \cdot R_{HS}/16 \parallel 31 \cdot R_{HS}/8 \parallel 31 \cdot R_{HS}/4 \parallel 31 \cdot R_{HS}/2 \parallel 31 \cdot R_{HS}$) = $R_{HS}$, and all parasitic capacitance contributed by the transistors, resistors, and MRM, i.e., a roughly constant and data-independent capacitance $C_{HS\text{-}DAC} + C_{MRM} \approx 30$ fF [13]. For a 2-GSym/s, 4-bit, 2.4-V DR data, the ($R_{HS}$·30-fF) time-constant should be at least less than 58 ps so that the HS-DAC output can spend less than 200 ps settling to its static analog-voltage level within an LSB/2 of its 4-bit resolution, i.e., exp(–200-ps/58-ps) < $1/(2^{4+1})$. Therefore, $R_{HS}$ needs to be less than 58-ps/30-fF ≈ 2 kΩ, which sets the static power consumption per HS-DAC discussed in the next paragraph. Overall, the O/E latency spends 300 ps (= 100-ps FO4 logic delays + 200-ps HS-DAC output settling time) out of the 500-ps clock-period budget (i.e., 2-GHz clock-rate).

The power consumption of each HS-DAC is contributed by both dynamic power and static power. First, the FO4 digital logics in each HS-DAC, including the DFFs, calibration encoder, data buffers, and push-pull data latches, mainly consume 0.2-mW dynamic power at 2-GSamp/s, which follows $C_L \cdot V_{DD}^2 \cdot f_{CLK}/2$ or $C_L \cdot (V_{DDH} - V_{DD})^2 \cdot f_{CLK}/2$ with the assumption of 50% Logic-1 and 50% Logic-0 data pattern per logic gate across multiple MVM operation cycles. Second, the static power consumption is mainly due to the DC current path from $V_{DDH}$ to GND of the resistor-divider formed by the parallel series-resistors and voltage-mode drivers as mentioned. Though the AC resistance of this HS-DAC architecture is constant, the DC path resistance or current between the $V_{DDH}$ and GND is data-dependent. By assuming the data patterns of $z_i$ with corresponding voltage-levels are uniformly distributed between 0 to ($2^B - 1$) across multiple MVM operation cycles, the average static power consumption per HS-DAC can be expressed as follows:

$$P_{HS\text{-}DAC,ST} = \frac{V_{DDH}^2}{R_{HS}} \cdot \frac{\sum_{k=1}^{2^{B-1}} (k-1) \cdot (2^B - k)}{2^{B-1} \cdot (2^B - 1)^2} \tag{7}$$

In the case of Fig. 2, where B = 4, $R_{HS} \approx 2$ kΩ, $V_{DDH} = 2.4$ V, each HS-DAC consumes 0.45-mW static power. Overall, the average power consumption of each HS-DAC is 0.65 mW, including both dynamic and static power.

The silicon area of each HS-DAC shown in Fig. 2 is around 50-um×20-um, including the voltage-mode driver, push-pull data latches, metal-finger capacitors for the AC-coupling, unsilicided poly-resistors for the voltage-divider, calibration encoder logics, and a static logic for resetting the data-dependent initial conditions of the push-pull data latches [9]. The 20-μm height per HS-DAC tile is designed to match the height per V-MRM tile.

### *C. Low-Power R2R Digital-to-Analog Data Converters*

The approach to realizing multiplication of each vector and matrix elements $y_{ij} \cdot z_j$ (e.g., $y_{11} \cdot z_1$ in Fig. 2) is to provide a transmission power-gain ∝ $y_{ij}$ on the top of the $\lambda_j$ light-wave power pre-modulated to ∝ $z_j/d$ in the i-th waveguide. That is, after the E/O conversion, O-MUX, and O-PS for $z_i$ becoming $z_j/d$ in the i-th MVM row, the element-to-element "multiplication" is basically done by another light-wave power modulation through the M-MRM$_{ij}$ only effective to the $\lambda_j$ light-wave power. Thus, a DAC is required to convert a digital multiplicand $y_{ij}$ to its corresponding voltage-level for setting the power-gain of M-MRM$_{ij}$, which is the same D/A and E/O operations in Section IV.B. However, the DAC speed requirement for converting $y_{ij}$ is not critical at all because the value or state of the matrix $Y_{d \times d}$ is pre-determined and static during the regular MVM operations as described in Section III for the attention-head computations. This fact beneficially allows to use low-speed but low-power small-area R2R voltage-divider architecture [29] to implement $d^2$ of DACs in a high-dimensional MVM accelerator as shown in Fig. 1.

Again, to accommodate a reasonable E/O DR, the non-linear transmission power-gain induced by each M-MRM across all possible $2^B$ voltage-levels (e.g., $2^4 = 16$ levels in Fig. 2) is alleviated by adding one calibration bit, and thus a digital calibration encoder is required for each R2R-DAC$_{ij}$ mapping the incoming B-bit $y_{ij}$ data to (B+1)-bit $y_{ij}$' data to perform better E/O conversion linearity while maintaining the same amount of $2^B$ voltage-levels (not $2^{B+1}$). The R2R-DAC$_{ij}$ shown in Fig. 2 is constructed by alternating series-R and parallel-2R resistances with push-pull switches to form a voltage-divider driving the capacitive electrode of M-MRM$_{ij}$ based on the value of $y_{ij}$. According to the number of binary bits of $y_{ij}$, an R2R-DAC can minimize the amount of the resistor-and-transistor components for generating all required voltage-levels, which is very beneficial to both power and area savings by taking advantage of static operations or low-speed requirements.

The concept of each parallel-2R segment in the R2R-DAC is similar to the bit-segment in an HS-DAC; both requires a voltage-mode driver architecture using push-pull switching transistors to short $2 \cdot R_U$ resistor to $V_{DDH}$ or GND. However, because of the low-speed R2R-DAC operation, the pull-up transistor can be driven by a two-stage static level-shifter as shown in the top-right of Fig. 2 to simultaneously maximize the E/O DR up to 2.4 V and maintain 45-nm CMOS reliability in 45SPCLO with negligible power consumption since all logics are under static states during the MVM operations. The only noteworthy power consumption is static power in the voltage-divider formed by the R2R network conducting a static current from $V_{DDH}$ to GND. Before showing the expression of average static power consumption per R2R-DAC, multiple indexes and variables need to be defined: "i" and "j" are still the row and column indexes of the matrix; by treating all R2R-DAC$_{ij}$ are identical, "i" and "j" do not involve in the power calculation actually; k = 1 ~ $2^B$ represents the voltage-level index; p = 1 ~ B represents the circuit-node index, q = 1 ~ B represents the Kirchhoff's Voltage Law (KVL) superposition index of each circuit-node; $R_U$ is the unit-resistor shown in Fig. 2; $R_p = (G_p/H_p) \cdot R_U$ is the one-side equivalent resistance of each circuit-node; $G_p$ and $H_p$ are integers; $G_p/H_p$ is a simplest fraction; $V_{k,p}$ is the KVL superposition voltage of each circuit-node and each voltage-level. By assuming the data patterns of $y_{ij}$ and corresponding voltage-levels are uniformly distributed between 0 to ($2^B$ – 1) across the entire matrix $Y_{d \times d}$, the average static power consumption per R2R-DAC can be expressed as follows:

$$G_1 = 1, H_1 = 0 \Rightarrow R_1 = \infty, \qquad p = 1$$

$$R_p = R_{p-1} || (2R_U) + R_U = \frac{G_p}{H_p} \cdot R_U, \qquad p = 2 \sim B$$

$$\frac{V_{k,p}}{V_{DDH}} = \sum_{q=1}^{B} y_{ij,k} \langle B - q \rangle \cdot \frac{\min\,[G_p, G_q]}{2^{p+q-1}}, \qquad \begin{matrix} k = 1 \sim 2^B \\ p = 1 \sim B \end{matrix}$$

$$P_{R2R\text{-}DAC,ST} = \frac{V_{DDH}^2}{R_U} \cdot \frac{\sum_{k=1}^{2^B} \sum_{p=1}^{B} y_{ij,k} \langle B - p \rangle \cdot \left(1 - \frac{V_{k,p}}{V_{DDH}}\right)}{2^{B+1}} \tag{8}$$

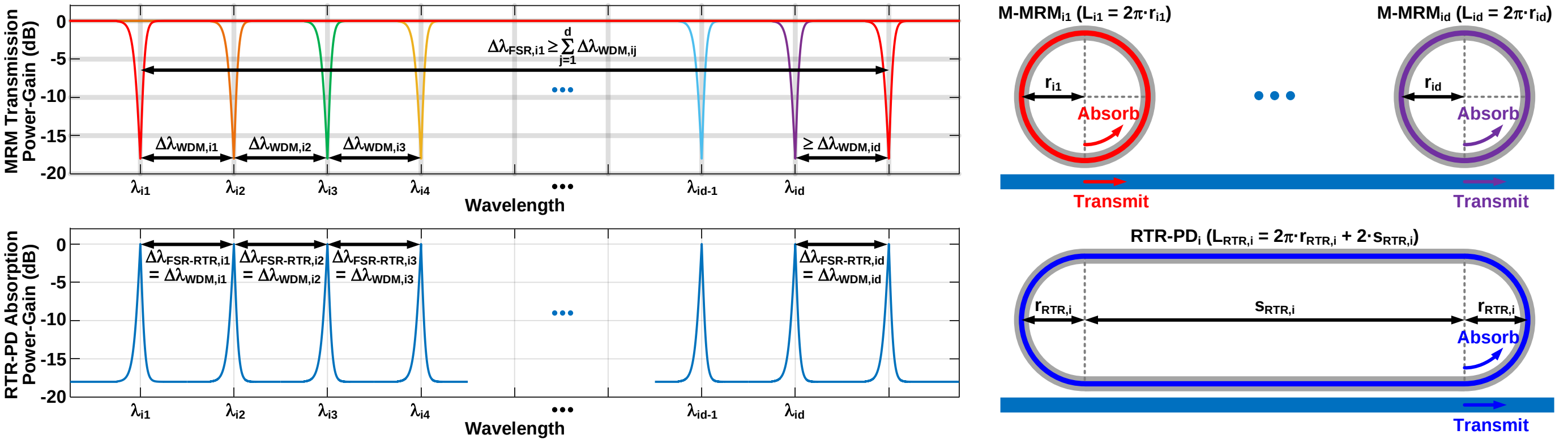

(a) (b)

Fig. 3. (a) The transmission responses of M-MRM$_{ij}$, j = 1 ~ d (or V-MRM$_i$, i = 1 ~ d, if $\lambda_{ij}$ is replaced with $\lambda_i$) and the absorption responses of RTR-PD$_i$. (b) The conceptual geometry of M-MRM$_{ij}$ and RTR-PD$_i$.

TABLE I
DESIGN TRADEOFF EXAMPLES AMONG MVM DIMENSIONS, WDM CROSSTALK, AND FABRICATION ERRORS

| d | $n_{eff}(\lambda_{ij})$ j = 1 ~ d | $n_g(\lambda_{ij})$ j = 1 ~ d | $m_{RTR,ij}$ j = 1 ~ d | $L_{RTR,i}$ | $\lambda_{ij}$ j = 1 ~ d | $\Delta\lambda_{WDM,ij}$ j = 1 ~ d | $m_{ij}$ j = 1 ~ d | $r_{ij}$ j = 1 ~ d | $\Delta r_{rms}/r_{ij}$ Error Scale |
|---|---|---|---|---|---|---|---|---|---|
| 32 | 3.74 ~ 3.73 | 5.02 ~ 4.98 | 2321 ~ 2290 | 951.32 µm | 1534.5 ~ 1550 nm | 0.49 ~ 0.51 nm | 71 or 72 | 4.63 ~ 4.76 µm | 1× |
| 32 | 3.76 ~ 3.73 | 5.06 ~ 4.98 | 1160 ~ 1129 | 469.01 µm | 1519.0 ~ 1550 nm | 0.97 ~ 1.03 nm | 35 or 36 | 2.25 ~ 2.38 µm | 2× |
| 16 | 3.74 ~ 3.73 | 5.02 ~ 4.98 | 1160 ~ 1145 | 475.66 µm | 1535.0 ~ 1550 nm | 0.99 ~ 1.01 nm | 71 or 72 | 4.63 ~ 4.76 µm | 1× |

In the case of Fig. 2, where B = 4, $R_U \approx 5$ MΩ, $V_{DDH}$ = 2.4 V, each R2R-DAC consumes about 7.2-µW. To reach up to about 5 MΩ with a small amount silicon area for each $R_U$, the resistance template is implemented by multi-stacked sub-threshold-region transistors [30] as the example shown in the top-right of Fig. 2, where $2 \cdot R_U$ is formed by four stacked P/N parallelized diode-connected transistors. Since the relative resistance ratios in the R2R network is more critical rather than the absolute resistance values, the process-corner variations and temperature coefficients of the active resistors are tolerable with proper transistor sizing under the 4-bit accuracy requirements. However, the nonlinearity due to data-dependent terminal voltages across the active resistors require extra attention. E.g., the resistance ($2 \cdot R_U$) between $V_{x,4}$ and $V_{k,4}$ in Fig. 2 is data-dependent since $V_{x,4}$ can be either $V_{DDH}$ or GND, and $V_{k,4}$ can vary from GND to (85/128)·$V_{DDH}$ based on the $V_{k,p}$ formula in Eq. (8). This paper proposed complimentary sub-threshold-region P/N active-resistor architecture: when $V_{x,4} = V_{DDH} > V_{k,4}$, the sub-threshold bias conditions are $0 < |V_{GS,PMOS}| < V_{th,PMOS}$ and $V_{GS,NMOS} = 0$; when $V_{x,4} = \text{GND} < V_{k,4}$, the sub-threshold bias conditions are $0 < |V_{GS,NMOS}| < V_{th,NMOS}$ and $V_{GS,PMOS} = 0$. That is, under these two possible $V_{x,4}$ conditions, the PMOS and NMOS sub-threshold biases are automatically set to be complimentary, which can cancel the first-order resistance nonlinearity due to the data-dependent $V_{x,4}$. On the other hand, the resistance nonlinearity due to $2^B$ different $V_{k,p}$ values based on $y_{ij}$<3:0> in general can be calibrated by the additional bit in $y_{ij}$'<3:-1> together with the M-MRM nonlinear power-gain calibration discussed in Section IV.D.

Though the transistor numbers and resistance values in the R2R-DAC all seem higher than those in the HS-DAC, the silicon area of each R2R-DAC shown in Fig. 2 is around 20-µm×10-µm, including the voltage-mode driver, push-pull data latches, sub-threshold-region active resistors, and calibration encoder logics. The 20-µm width per R2R-DAC tile is designed to match the width per M-MRM tile.

### D. *Micro-Ring Modulators & E/O Conversions*

Once a data element ($z_i$ or $y_{ij}$) of the vector or matrix is converted and settled to an analog-voltage level through either an HS-DAC or R2R-DAC at the P-type electrode of its V-MRM or M-MRM, the light-wave power of a certain wavelength, which is located within the MRM resonance bandwidth, can be effectively modulated according to the voltage delta between the N-type (i.e., the DC voltage from $V_{B,MRM}$) and P-type (i.e., the settled DC voltage from the DAC output) electrodes of the MRM (i.e., the reversed bias between the MRM P/N junction).

The resonance wavelength, coupling strength, optical loss, and quality factor of an MRM are mainly determined by its transmission waveguide widths, ring waveguide width, transmission-to-ring waveguide gap, and especially ring radius, which is the dominant factor in setting the footprint of the MRM. The radius $r_{ij}$ design of M-MRM$_{ij}$ (or $r_i$ of V-MRM$_i$) in the whole WDM spectrum shall at least satisfy two criteria. First, the minimum free spectral range (FSR) is confined by all WDM isolation spacing $\Delta\lambda_{WDM,ij}$ between adjacent $\lambda_{ij}$ and the number of $\lambda_{ij}$ for the WDM (i.e., "d") [31] as shown in the top of Fig. 3(a). Second, each M-MRM$_{ij}$ resonance wavelength $\lambda_{ij}$ under a particular resonance mode $m_{ij}$ (an integer) is corresponding to its effective refractive index $n_{eff}(\lambda_{ij})$, silicon propagation constant $\beta(\lambda_{ij})$, and ring circumference $L_{ij} = 2\pi \cdot r_{ij}$; in other words, when the resonance condition is satisfied in the MRM cavity (i.e., $L_{ij}$), a constructive interference is established by a certain wavelength having its round-trip phase shift equal to an integer multiple of $2\pi$ [32]. These two criteria are summarized in Eq. (9) and (10), respectively, as follows:

$$\Delta\lambda_{FSR,ij} = \frac{\lambda_{ij}^2}{n_g(\lambda_{ij}) \cdot 2\pi \cdot r_{ij}} \geq \sum_{j=1}^{d} \Delta\lambda_{WDM,ij}$$
$$\Rightarrow r_{ij} \leq \frac{\lambda_{ij}^2}{n_g(\lambda_{ij}) \cdot 2\pi \cdot \sum_{j=1}^{d} \Delta\lambda_{WDM,ij}} \quad (9)$$

$$2\pi \cdot m_{ij} = \beta(\lambda_{ij}) \cdot L_{ij} = \frac{2\pi}{\lambda_{ij}} \cdot n_{eff}(\lambda_{ij}) \cdot 2\pi \cdot r_{ij}$$
$$\Rightarrow r_{ij} = \frac{\lambda_{ij}}{2\pi \cdot n_{eff}(\lambda_{ij})} \cdot m_{ij} \quad (10)$$

where $\Delta\lambda_{FSR,ij}$ is the FSR of M-MRM$_{ij}$; $n_g(\lambda_{ij})$ is the silicon group index; $2\pi/\lambda_{ij}$ is the free-space propagation constant; $\lambda_{ij}$ presents the free-space resonance wavelength of M-MRM$_{ij}$ although the light-waves propagate in the silicon. To describe the M-MRM$_{ij}$ (or V-MRM$_i$) design procedure, a practical example with realistic parameter values for determining the radius $r_{ij}$ of M-MRM$_{ij}$ is demonstrated in the following paragraphs as the initial design without considering any post-process trimming or thermal control.

Typically, the nominal wavelength-spectrum bandwidth in the silicon of 45SPCLO is in the range of 1180 nm to 1550 nm [12]. Therefore, with the MRM quality-factor (i.e., Q) up to 10,000 [12] in 45SPCLO, the WDM isolation spacing $\Delta\lambda_{WDM,ij}$ across 32 (= d) wavelengths is initially set to ~0.5 nm, and the maximum WDM wavelength $\lambda_{i32}$ is set to 1550 nm in this example. At this point, it is important to note that the exact 32 resonance wavelengths ($\lambda_{ij}$, j = 1 ~ 32) distributed from 1534.5 nm to 1550 nm and 32 isolation spacing ($\Delta\lambda_{WDM,ij}$, j = 1 ~ 32) slightly varying from 0.49 nm to 0.51 nm listed in Table I are actually determined by the design equation of the optical comb-based RTR-PD discussed in Section IV.E.

With all designated $\lambda_{ij}$, $\Delta\lambda_{WDM,ij}$, and their corresponding $n_g(\lambda_{ij})$, the FSRs of all M-MRM$_{ij}$ have to meet the criterion of $\Delta\lambda_{FSR,ij} \geq 32 \cdot 0.5$ nm = 16 nm, and then the upper bound of each $r_{ij}$ can be specified based on Eq. (9). Finally, by properly choosing the mode integer $m_{ij}$ with corresponding $n_{eff}(\lambda_{ij})$, all $r_{ij}$ of M-MRM$_{ij}$ can be obtained and distributed from 4.63 μm to 4.76 μm in this example. By following the same procedure and setting different values of "d" and initial $\Delta\lambda_{WDM,ij}$, the design tradeoffs among the MVM dimension, WDM crosstalk, and MRM fabrication error are summarized in Table I, including three different scenarios: the case one has the worst WDM crosstalk due to the smallest $\Delta\lambda_{WDM,ij}$; the case two has the worst MRM fabrication error $\propto (2\pi \cdot \Delta r_{rms})/(2\pi \cdot r_{ij})$ where $2\pi \cdot \Delta r_{rms}$ is the circumference error of M-MRM$_{ij}$ due to random process variation and independent from $r_{ij}$; the case three has the worst computation throughput because of the smallest values of "d" out of these three scenarios. In any case, the silicon area per MRM can be confined within a 20-μm×20-μm tile, including the central ring area and peripheral keep-out halo.

The linearity of the entire analog circuit, photonic dynamic range, and signal conversions in the MPE accelerator is extremely crucial; this issue could continuously dominate the development of photonic computing. Specifically in the E/O conversions, the deterministic nonlinearity is mainly caused by the sigmoid-like high-Q power transmission response shown in Fig. 4(a) as the zoom-in version of Fig. 3(a) for M-MRM$_{11}$ simulated by the approach of Verilog-A modeling [33], [34] even though the transmission resonance can be almost linearly shifted with the reverse bias driven by a linear DAC.

For the example in Fig. 4(a), the wavelength of $\lambda_1$-laser carrying data information in its light-wave power is designated at 1534.5 nm to match the resonance wavelength of M-MRM$_{11}$ with the reverse bias at 0 V corresponding to $y_{11}$<3:0> = <0000>. Under 16 different reverse biases generated by a 4-bit linear DAC according to $y_{11}$ varying from <0000> to <1111>, the MRM resonance wavelength and transmission response are shifted horizontally in a constant rate of 0.04-nm/1-V. However, this causes the power attenuation (i.e., power gain < 1) of $\lambda_1$-laser nonlinearly distributed across the vertical E/O DR as highlighted by black triangular marks in Fig. 4(a) and 4(b) on dB scale. The same $y_{11}$ vs. E/O conversion curves are also shown in Fig. 4(e) but on a linear scale for the sake of linearity demonstration. To alleviate this nonlinear effect, an additional calibration bit is required in the DAC to generate 32 reverse-bias options so that the encoder and calibration logic can pick 16 out of 32 reverse biases to fit the linear E/O transfer curve as highlighted by blue circle marks in Fig. 4(e). In other words, $y_{11}$<3:0> can be nonlinearly 1-to-1 mapped to $y_{11}$'<3:-1> to cancel the nonlinearity and eventually obtain better linearity in the E/O conversion. Therefore, although the transmission response shifts nonlinearly according to $y_{11}$<3:0> (mapped to

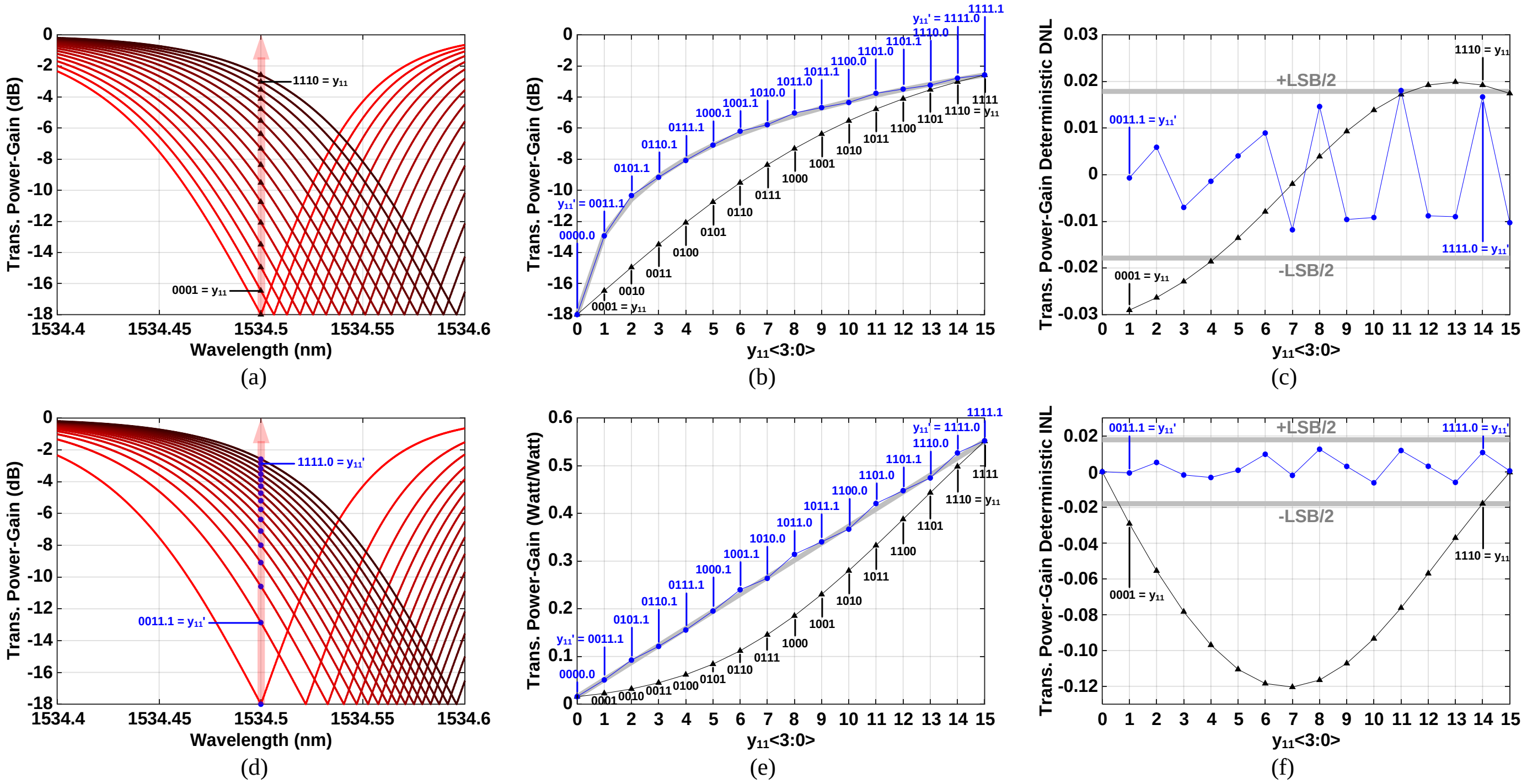

Fig. 4. (a) The transmission responses of M-MRM$_{11}$ modulated by a 4-bit linear DAC. (b) The transfer curves of $y_{11}$<3:0> vs. 4-bit nonlinear (black) and linearized (blue) E/O conversions on a dB scale. (c) The deterministic DNL of the 4-bit nonlinear (black) and linearized (blue) E/O conversions. (d) The transmission responses of M-MRM$_{11}$ modulated by a 4-bit nonlinear DAC for the linearization. (e) The identical transfer curves shown in (b) but on a linear scale. (f) The deterministic INL of the 4-bit nonlinear (black) and linearized (blue) E/O conversions.

16 out of 32 of $y_{11}$'<3:-1>) in Fig. 4(d), the power attenuation of $\lambda_1$-laser relatively has better linearity, which is very essential to the linear-algebra accelerator. The linearity enhancement of this technique is quantified by the deterministic E/O differential nonlinearity (DNL) and integral nonlinearity (INL) in Fig. 4(c) and 4(f), respectively. Both are reduced down to the range within ±LSB/2 of the 4-bit E/O conversion process.

### *E. Racetrack-Resonator Photodetectors & O/E Conversions*

The zero change to the underlying CMOS process in the monolithic SiPh IC fabrication technology brings tremendous integration and energy efficiency for optical communication and computation systems. However, one of the major downsides is the performance degradation of photonic characteristics. E.g., the responsivity of a 50-µm×5-µm linear PD could be limited to about 0.023 A/W [14] in 45SPCLO. A clever solution relies on the absorption property of a resonator to enhance the responsivity of a micro-ring-resonator based PD up to 0.55 A/W [13], [35], but this is only beneficial to the WDM receivers requiring to extract the power of individual wavelengths from an incoherent communication channel.

To simultaneously enhance the PD responsivity and detect the aggregate light-wave power across all WDM wavelengths used in the MPE-MVM accelerator, this paper proposed an optical comb-based racetrack-resonator (RTR) [36] PD as shown in Fig. 1, 2, and 3. The fundamentals of racetrack and micro-ring resonators are basically identical, but the dimensions of their resonance cavity are quite different. As shown in Fig. 3(b), if the waveguide widths and gaps of the micro-ring and racetrack resonators are unified, then their resonance cavity lengths, $L_{ij}$ and $L_{RTR,i}$, are the primary design parameters determining their power absorption responses. Note that both power transmission and detection both rely on the concept of a resonance cavity, but their signal observation points are different. In power transmissions, an MRM is set to absorb a certain amount of light-wave power of a certain wavelength so that only the residual light-wave power of this wavelength can pass through or be transmitted in the straight waveguide. In power detections, an RTR-PD is set to absorb all light-wave powers of all wavelengths and convert the aggregate optical power to a photocurrent in the electronic domain; this O/E conversion efficiency is quantified by the responsivity (A/W). For the example shown in the bottom of Fig. 3(b), under the resonance condition, the RTR cavity simultaneously establishes constructive inferences with all $\lambda_{ij}$ wavelengths so that each MVM dot-project result carried by all $\lambda_{ij}$ light-wave powers in each straight waveguide (i.e., each MVM row) can be altogether absorbed in the RTR cavity, which can improve the photodetection responsivity by utilizing the high-Q property of the RTR on all $\lambda_{ij}$. That is, on the condition of critical coupling across all $\lambda_{ij}$, the aggregate optical power received by RTR-PD$_i$ is the linear summation of all $\lambda_{ij}$ power absorptions [13] through the concurrent multi-wavelength resonances in the common racetrack cavity, which is the key idea of responsivity enhancement for the broadband photodetection as follows:

$$P_{Absorb\text{-}RTR,i} \approx \frac{Q_{RTR,i}}{\pi \cdot L_{RTR,i}} \cdot \sum_{j=1}^{d} \frac{\lambda_{ij} \cdot P_{trans\text{-}MRM,ij}(y_{ij})}{n_g(\lambda_{ij})} \quad (11)$$

To design an RTR-PD having a consistent power absorption gain for all WDM laser wavelengths as shown in the bottom of Fig. 3(a), the perimeter $L_{RTR,i}$ of RTR-PD$_i$ for the whole WDM spectrum shall at least satisfy two criteria. First, the FSR $\Delta\lambda_{FSR\text{-}RTRij}$ of each resonance wavelength $\lambda_{ij}$ determines each WDM isolation spacing $\Delta\lambda_{WDM,ij}$ [31]. Second, each resonance wavelength $\lambda_{ij}$ under a particular resonance mode $m_{RTR,ij}$ (an integer) is corresponding to its effective refractive index $n_{eff}(\lambda_{ij})$, silicon propagation constant $\beta(\lambda_{ij})$, and the common perimeter $L_{RTR,i} = 2\pi\cdot r_{RTR,i} + 2\cdot s_{RTR,i}$ [32]. These two criteria are summarized in Eq. (12) and (13), respectively, as follows:

$$\Delta\lambda_{FSR\text{-}RTR,ij} = \frac{\lambda_{ij}^2}{n_g(\lambda_{ij})\cdot L_{RTR,ij}} = \Delta\lambda_{WDM,ij}$$
$$\Rightarrow L_{RTR,i} = \frac{\lambda_{ij}^2}{n_g(\lambda_{ij})\cdot \Delta\lambda_{WDM,ij}} \quad (12)$$

$$2\pi\cdot m_{RTR,ij} = \beta(\lambda_{ij})\cdot L_{RTR,i} = \frac{2\pi}{\lambda_{ij}}\cdot n_{eff}(\lambda_{ij})\cdot L_{RTR,i}$$
$$\Rightarrow L_{RTR,i} = \frac{\lambda_{ij}}{n_{eff}(\lambda_{ij})}\cdot m_{RTR,ij} \quad (13)$$

To satisfy both Eq. (12) and (13), all resonance wavelengths $\lambda_{ij}$, j = 1 ~ d, of the entire WDM spectrum are determined by the resonance mode requirements of RTR-PD$_i$ as shown as follows:

$$m_{RTR,ij} = \frac{n_{eff}(\lambda_{ij})}{n_g(\lambda_{ij})}\cdot\frac{\lambda_{ij}}{\Delta\lambda_{WDM,ij}}$$
$$= \frac{n_{eff}(\lambda_{ij})}{n_{eff}(\lambda_{ij}) - \lambda_{ij}\cdot\frac{\partial n_{eff}(\lambda_{ij})}{\partial\lambda_{ij}}}\cdot\frac{\lambda_{ij}}{\Delta\lambda_{WDM,ij}} \quad (14)$$

Note that each $m_{RTR,ij}$ has to be a unique integer for j = 1 ~ d since all constructive interferences simultaneously occur in the same cavity, $L_{RTR,i}$. In other words, the only way to set distinguishable "d" resonance wavelengths $\lambda_{ij}$ by Eq. (13) with a common $L_{RTR,i}$ is to assign "d" individual $m_{RTR,ij}$. In summary, the value of "d" with the requirements in Eq. (12), (13), and (14) determines the distribution of $\lambda_{ij}$ and $\Delta\lambda_{WDM,ij}$ of the entire WDM wavelength spectrum for the MVM operation; then the laser wavelengths and the radii of all V-MRM$_i$ and M-MRM$_{ij}$ described in Section IV.D shall be designed accordingly to match the power transmission and absorption spectrums as shown in Fig. 3(a).

By using the same example in Section IV.D with a few iterations, the consecutive integer $m_{RTR,ij}$, j = 1 ~ 32, are chosen from 2321 down to 2290 so that $\lambda_{ij}$ and $\Delta\lambda_{WDM,ij}$, j = 1 ~ 32, can be determined to satisfy the targeted ~0.5-nm WDM isolation spacing with $L_{RTR,i}$ = 951.32 μm as summarized in Table I. About the area of RTR-PD$_i$, if the radius $r_{RTR,i}$ of the left/right-end half-circles is set to 5 μm, then the length $s_{RTR,i}$ of the top/bottom straight waveguides is 459.95 μm. Therefore, including the primary racetrack-resonator and peripheral keep-out halo, the silicon area per RTR-PD is a 480-μm×20-μm tile.

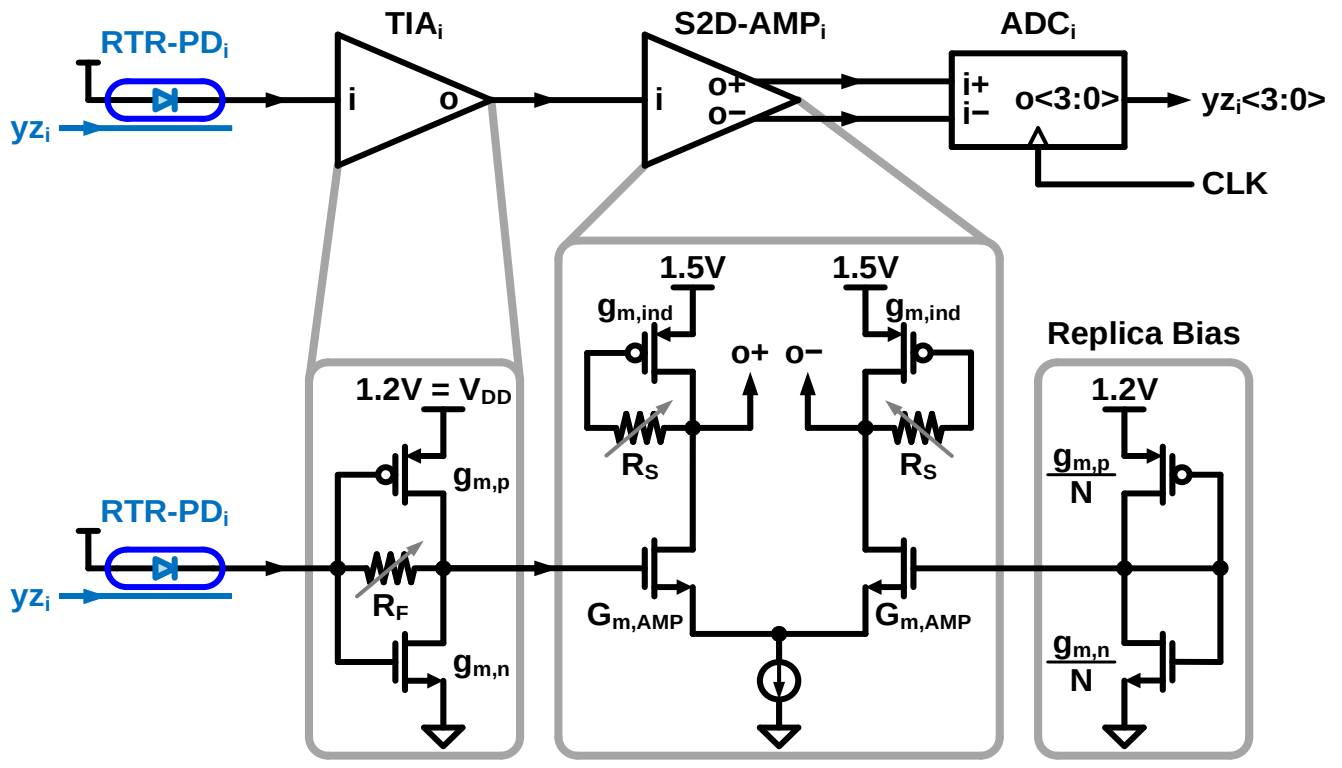

Fig. 5. The block diagram of each MPE-MVM dot-product O/E conversion circuit, and the schematics of TIA$_i$ and S2D-AMP$_i$.

### *F. Transimpedance Amplifiers, Single-Ended-to-Differential Amplifiers & Analog-to-Digital Data Converters*

Once each dot-product yz$_i$ converted from the aggregate light-wave power to a photocurrent in the electronic domain through RTR-PD$_i$, the following CMOS circuit blocks, including the TIA, single-end-to-differential amplifier (S2D-AMP), and ADC as shown in Fig. 5, further convert yz$_i$ from the analog photocurrent to a B-bit digital word. Up to this point, the entire signal flow of the MVM operation is basically completed in terms of the end-to-end digital data format, computation outcome, and operation clock-cycle.

The TIA in Fig. 5 is implemented by a voltage-to-current feedback-amplifier architecture. The feedforward path is formed by a complimentary P/N transconductance ($G_{m,TIA} = g_{m,p} + g_{m,n}$) stage while both share the single DC bias current from the TIA supply ($V_{DD}$) to GND. The feedback path is formed by a resistor $R_F$, which plays the key role of the TIA gain, bandwidth, and input/output impedance with $G_{m,TIA}$. In addition, this TIA is self-biased because of a zero DC feedback current through $R_F$ and series P/N diode-connection; the input and output DC bias voltages can be self-locked at the operating point of ($V_{GS,n} + V_{SG,p}$) = $V_{DD}$ without requiring other biasing mechanism and extra power consumption. This TIA architecture has been broadly used in high-speed optical receiver front-ends [11], [37] because of its simplicity for both high-bandwidth and low-power characteristics. The simplified output resistance, transfer function, and average input-referred noise power spectrum density (PSD) [37] of this TIA architecture are shown in Eq. (15), (16), and (17), respectively:

$$R_{TIA} \approx \frac{R_F}{1 + G_{m,TIA}\cdot R_F} \quad (15)$$

$$TF_{TIA}(s) \approx -\frac{G_{m,TIA}\cdot R_F\cdot R_{TIA}}{1 + s\cdot R_{TIA}\cdot C_{TIA}} \quad (16)$$

$$\overline{I_{n,in,TIA}^2} \approx \frac{2\pi\cdot\kappa\cdot T}{R_F^2}\cdot\left(\frac{\gamma}{G_{m,TIA}} + \frac{1}{G_{m,TIA}^2\cdot R_{TIA}}\right) \quad (17)$$

where κ = 1.38×10$^{-23}$ J/K is the Boltzmann constant; T = 300 K is the thermal dynamic temperature in Kelvin; γ ≈ 2.5 is the excess noise coefficient for deep submicron technology; $C_{TIA}$ ≈ 30 fF is the capacitive load at the TIA$_i$ output, which is mainly contributed by the input capacitance of S2D-AMP$_i$. Based on the design example so far with the 0.5-A/W RTR-PD$_i$ responsivity and 670-μW DR of the aggregate WDM light-wave power, the DR of the TIA input photocurrent is 335 μA. To achieve 335-mV DR at the TIA output, $G_{m,TIA}$ and $R_F$ are chosen to be 1 mA/V and 1.65 kΩ, respectively, so that the TIA DC gain can be about 335-mV/335-μA = 1 kΩ ≈ $|TF_{TIA}(0)|$ based on Eq. (16). Meanwhile, the TIA bandwidth ≈ $1/(2\pi \cdot R_{TIA} \cdot C_{TIA})$ reaches 8.52 GHz, which is sufficiently higher than the 1-GHz Nyquist frequency of the 2-GSym/s per $yz_i$ data. More importantly, this design choice leads to the average input-referred noise PSD of this TIA as the first active-stage of the O/E interface is as low as 6.26 pA/$\sqrt{\text{Hz}}$ based on Eq. (17). About the static power consumption, the TIA for 1-mV/A $G_{m,TIA}$ and its scaled replica for generating identical DC common-mode voltages for the S2D-AMP$_i$ differential-pair together consume about 0.1 mW from the 1.2-V supply.

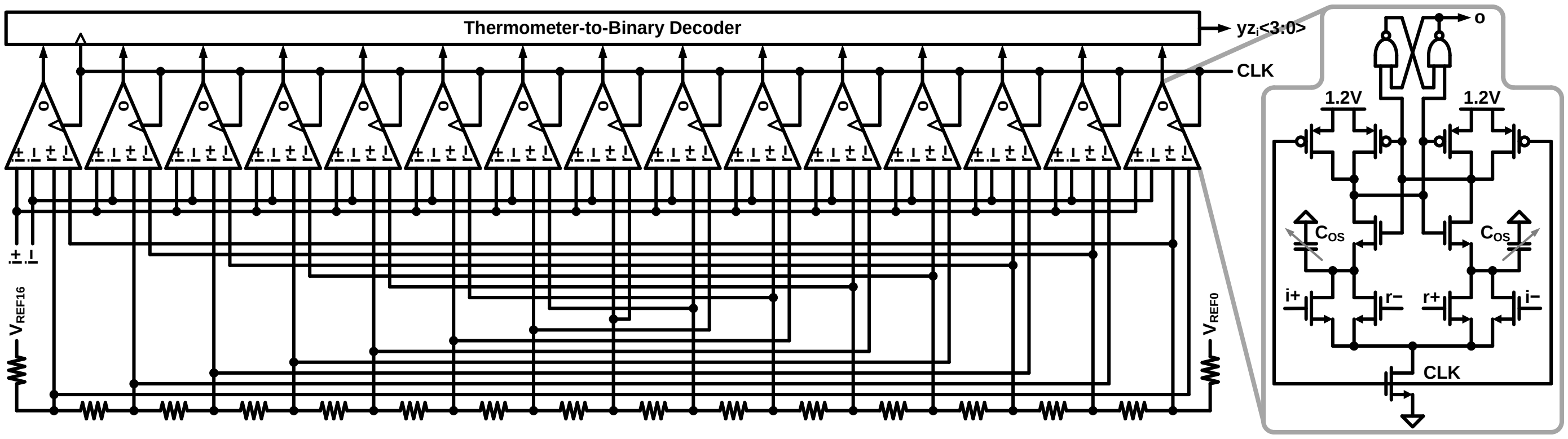

Fig. 6. The schematic of $ADC_i$ in each MPE-MVM dot-product O/E conversion circuit, and the schematic of the SAL-based clocked comparator.

The second-stage, S2D-AMP$_i$, is implemented by a fundamental common-source differential amplifier with a pair of active-inductor loads to mainly convert and buffer the single-ended TIA output to a differential signal for minimizing asymmetrical kickbacks and common-mode noise at the input of the following $ADC_i$ as shown in Fig. 5. The active-inductor load [38] is formed by a diode-connected PMOS with a tunable feedback resistor $R_S$ to enhance the high-frequency gain for sharpening data symbol transitions and to intentionally unbalance the gains of the positive and negative outputs for cancelling the nonideal single-ended-to-differential conversion process due to the finite output impedance of the tail current source. Based on the same design example so far, if the S2D-AMP$_i$ output capacitive load $C_{AMP}$ mainly contributed by the input capacitance of $ADC_i$ is about 80 fF, its output resistance $R_{AMP}$ is chosen to be 600 Ω so that the default output 3-dB bandwidth ≈ $1/(2\pi \cdot R_{AMP} \cdot C_{AMP})$ is about 3.32 GHz, which dominates speed of the entire MVM operation. Fortunately, without extra power consumption, the active-inductor load can effectively enhance the bandwidth up to 5.5 GHz, which is sufficient for the 1-GHz Nyquist frequency of the 2-GSym/s per 4-bit $yz_i$ data. This is a good example to remind that, in the monolithic integrated MVM, the S2D-AMP$_i$ output (or $ADC_i$ input) is usually the critical bandwidth node (e.g., 5.5 GHz) of the entire WDM signal path rather than the TIA output bandwidth (e.g., 8.52 GHz) in general. On the other hand, in other heterogeneously integrated MVMs, the critical bandwidth node could be at the TIA input, which is the interface between the separate electronic and photonic ICs requiring dedicated I/O bumps, ESDs, and passive-inductor peaking (T-coil) circuits.

The DC voltage-gain $|TF_{AMP}(0)| = G_{m,AMP} \cdot R_{AMP}$ of S2D-AMP$_i$ is set to 3×, so that the 335-mV TIA output DR can be converted to 1-$V_{diff}$ ADC input DR through S2D-AMP$_i$. Also, $G_{m,AMP}$ and the static power consumption of S2D-AMP$_i$ can be respectively determined as 5 mA/V and 0.75 mW from the 1.5-V supply.

The third-stage, $ADC_i$, is implemented by a B-bit flash ADC architecture consisting of ($2^B$ – 1) strong-arm-latch (SAL) based clocked-comparators [39] as shown in Fig. 6. The flash ADC architecture is suitable for high-speed and low-resolution applications with the downside of $2^B$ exponential-growth of the circuit area and dynamic power. In the case of this on-chip 4-bit O/E interface, the 15 SAL-based comparators per flash ADC actually offer adequate compromise between area and conversion rate (i.e., sampling rate = 2 GS/s) without consuming static power. Each SAL-based clocked comparator consists of a SAL followed by a RS-latch to form an edge sampled DFF as a full clock-cycle register. The dual differential pairs of each SAL are exploited to compare the volage difference between its differential-input and differential-reference voltages. The SAL architecture allows itself to complete the signal integration, regeneration, and decision within a half-period of the sampling clock by using single phase of the sampling clock and concept of integrating amplifications so that the average power consumption of 4-bit ADC is about 1.2 mW (= 15×80-μW). Note that additional 3-bit capacitor-banks ($C_{OS}$) for SAL offset/nonlinearity cancellations and a common resistor-ladder for all SALs differential reference-voltage generations both consume negligible power.

The limitation of the low-power O/E circuit designs mentioned above is bounded by the A/D accuracy criterion along with the specification of the laser injection power for the WDM-based MVM operation. Therefore, the design iteration is necessary to consider the noise power from the primary contributors in this O/E interface, including the RTR-PD$_i$ shot noise ($\overline{I^2_{n,PD}}$), TIA$_i$ circuit output noise ($\overline{V^2_{n,TIA}}$), and S2D-AMP$_i$ circuit output noise ($\overline{V^2_{n,AMP}}$). The overall noise power at the $ADC_i$ input can be expressed as follows:

$$P_{n,O/E} \approx \int_0^\infty \overline{I^2_{n,PD}}(f) \cdot |TF_{TIA}(f)|^2 \cdot |TF_{AMP}(f)|^2 \cdot df + \int_0^\infty \overline{V^2_{n,TIA}}(f) \cdot |TF_{AMP}(f)|^2 \cdot df + \int_0^\infty \overline{V^2_{n,AMP}}(f) \cdot df$$
$$\approx 0.5qI_{PD}\left(G^2_{m,TIA}R^2_F R^2_{TIA}\right)G^2_{m,AMP}R_{AMP}/C_{AMP} + kT\left(\gamma G_{m,TIA}R^2_{TIA} + R_{TIA}\right)G^2_{m,AMP}R_{AMP}/C_{AMP} + 2kT\left(\gamma G_{m,AMP}R_{AMP} + \gamma g_{m,ind}R_{AMP} + 1\right)/C_{AMP} \quad (18)$$

where q = 1.602×10$^{-19}$ C is the elementary charge; $g_{m,ind}$ is the transconductance of the PMOS used for the active-inductor load; $R_{AMP} \approx R_S/(1+g_{m,ind}\cdot R_S) \approx 600\ \Omega$ is the output resistance of S2D-AMP$_i$. In Eq. (18), the frequency-domain integrals are simplified by only considering the bandwidth of S2D-AMP$_i$ due to its bandwidth domination. Based on the circuit design parameters so far with the maximum $I_{PD} \approx 335$ µA, the overall noise power at the ADC$_i$ input is around 11 µW. Compared to the quantization noise of power of the 4-bit flash ADC = $V^2_{LSB}/12 = (1\text{-}V_{diff}/15)^2/12$ = 370 µW, there is still a margin of 2.5 bits ≈ 15.3 dB = 10·log$_{10}$(370-µW/11-µW) to accommodate dark noise, flicker noise, supply noise, clock jitter, resistor-ladder noise, comparator noise, and residual offset, which are not included in the noise estimation of Eq. (18).

### G. *Optical Power Splitters & Laser Injection Power*

The MPE-MVM accelerator as shown Fig. 1 requires WDM laser comb sources as being developed in [40] to interface with the optical gratings on the monolithic SiPh chip. After the E/O conversion for the individual elements of $Z_{d\times1}$, the on-chip O-MUX perform an analog summation by merging all independent $z_i$, i = 1 ~ d, into a single waveguide, and then the O-PS duplicates the pre-summed vector elements into "d" identical waveguides ready for the following WDM-based MVM operation. To realize the high fan-out O-PS, this paper exploited the concept of adiabatic Y-junction power splitters possessing low-loss, high-bandwidth, high-polarization insensitivity, and high-tolerance to fabrication errors by using a nonlinear taper coupling technique to shorten the horizontal dimension of each 50/50 Y-junction power splitter [41]. In the case of a fan-out of "d" WDM-based MVM, the O-PS contains log$_2$(d) stages horizontally, and each stage is vertically formed by multiple 50/50 Y-junction power splitters in parallel of a number from $2^0$ (= 1) to $2^{\log2(d)-1}$ (= d/2) for the first to last stages, respectively. The simulation result of the 50/50 Y junction power slitter in Fig. 7 on the 160-nm silicon-on-insulator (SOI) 45SPCLO platform shows that the 17.5-µm nonlinear taper coupler within a total 35-µm footprint, including the length from the single horizontal fan-in to two horizontal fan-outs, can reach transmission loss less than 0.07 dB, which is consistent with the result reported in [42], for the transverse electric (TE) mode in the range of 1530-nm to 1550-nm WDM bandwidth as specified in Table I. To match the 20-µm height per MVM row, the overall silicon area of a 1-to-d O-PS is [log$_2$(d)·35-µm]×[d·20-µm], which is 175-µm×640-µm

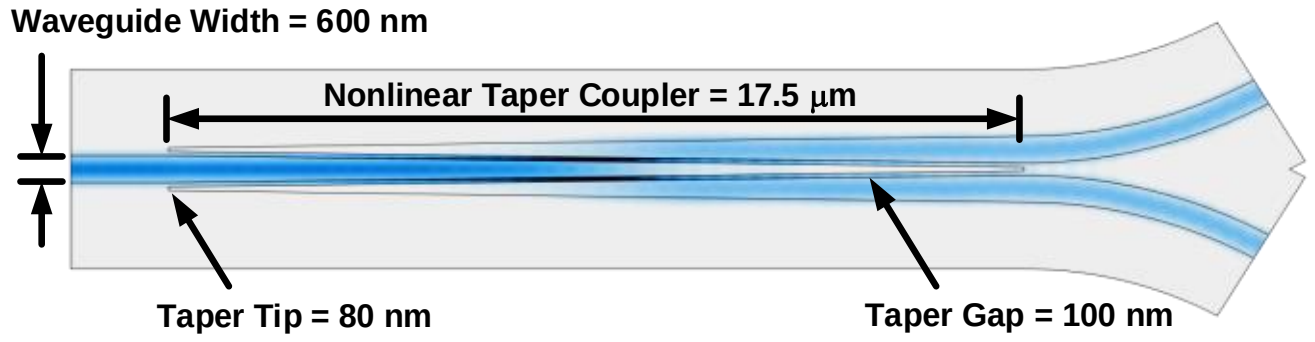

Fig. 7. The mask layout and dimensions of the 50/50 Y-junction power-splitter with the nonlinear taper coupling technique.

when d = 32.

The aggregate signal power loss "log$_2$(d)·(3.01 + 0.07)-dB" of a 1-to-d O-PS is consolidated with the transmission and absorption losses of MRMs and RTR-PD in each MVM row to estimate the required laser-injection power of each wavelength for satisfying the input DR and signal-to-noise ratio of O/E conversion as discussed so far. That is, each laser-injection power ($P_\lambda$) must confront log2(d) stages Y-junctions (3.01-dB power splitting and 0.07-dB loss per stage), two MRM transmission DR losses (2.5-dB loss per V-MRM$_i$ or M-MRM$_{ij}$ based on Fig. 4), and one RTR-PD absorption DR loss (2.5-dB loss per RTR-PD$_i$). Meanwhile, the aggregate absorption of all wavelength powers per RTR-PD needs to be confined within the linear DR of the O/E conversion circuit. This criterion for the laser-injection power per wavelength can be expressed as follows:

$$P_\lambda \cdot \left(10^{\frac{-2.5\text{-}dB}{10}}\right)^3 \cdot \left(\frac{1}{d} \cdot 10^{\frac{-\log_2(d)\cdot 0.07\text{-}dB}{10}}\right) \cdot d \le DR_{E/O} \quad (19)$$

where $P_\lambda$ is not a strong function of "d" because $DR_{E/O}$ stays constant regardless of "d", and only the number of the Y-junction stages would gradually increase the power loss within each MVM row. For the case of d = 32 and O/E DR = 670 µW, $P_\lambda$ is about 4.08 mW, and total 32 laser-injection power for this WDM-based MVM operation is 130.7 mW, which however is directly proportional to the number of WDM wavelengths "d".

### H. *Thermal Tuning & Post-Fabrication Trimming*

The high thermo-optic coefficient (1.86×10$^{-4}$ 1/K) of silicon [43] makes SiPh devices extremely sensitive to temperature variations; therefore, a proper thermo-control mechanism is necessary to stabilize the environment temperature for maintaining consistent characteristics of SiPh devices. On the other hand, this high temperature sensitivity also helps to enable high-resolution SiPh characteristic tunability beyond the achievable resolution of fabrication process especially for the WDM-based MVM; e.g., the 32 different radii of all M-MRM$_{ij}$ listed in Table I are distributed from 4.63 µm to 4.76 µm, which are impossible to be explicitly fabricated merely relying on the mask resolution of 45SPCLO. In CMOS compatible SiPh process technology, tungsten heaters have been widely used for thermo-optic tuning [44], and, according to [45], a tungsten heater power efficiency per MRM can reach about 2.4 mW/Δλ$_{FSR}$. Therefore, if a d-by-d MPE-MVM accelerator contains "d·(1 + d)" MRMs and "d" RTR-PDs within a 1-mm$^2$ silicon area, the tuning ranges of each MRM and RTR-PD need to cover about one $\Delta\lambda_{WDM} \approx \Delta\lambda_{FSR}/d \approx \Delta\lambda_{FSR,RTR}$, so the total heater power per d-by-d MVM is less than $d\cdot(1 + d)\cdot(2.4\text{-mW})/d + d\cdot 2.4\text{-mW}$, which is 156 mW when d = 32.

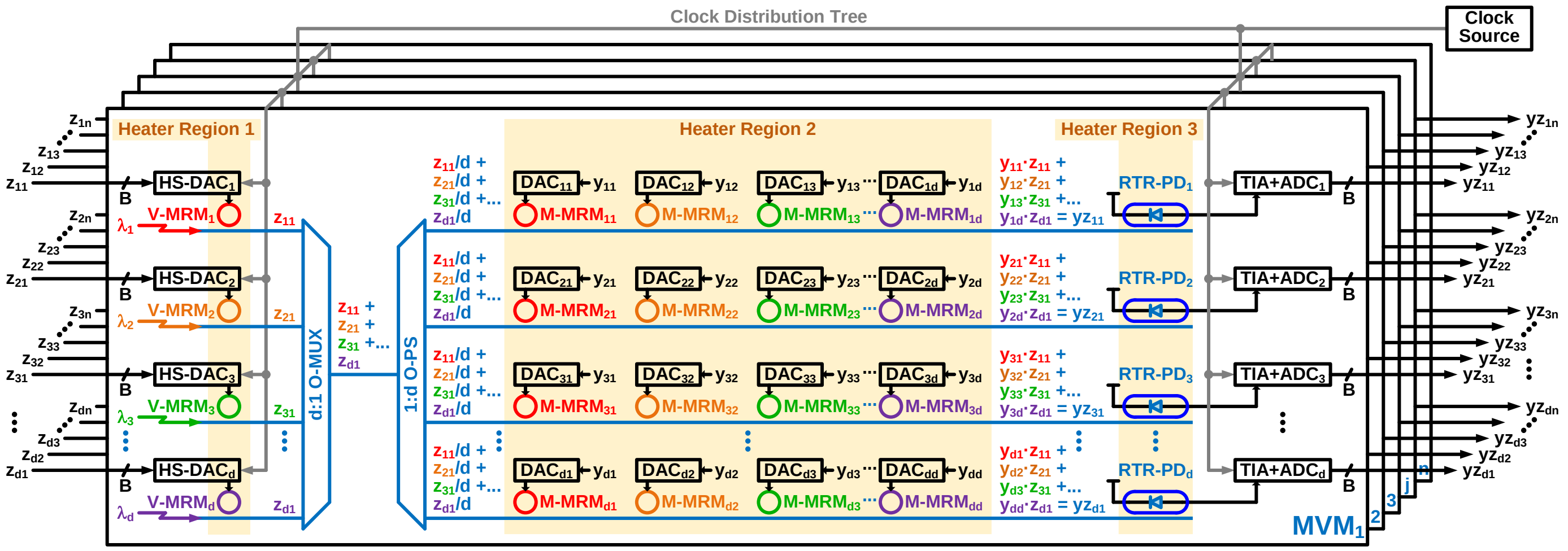

Fig. 8. The system block diagram of the MPE-MMM linear-algebra accelerator implemented by the MVM parallelism approach. Note that each "yz" presents a "single-word" variable, NOT "y times z".

Meanwhile, to accommodate a large number of MRMs for high-dimensional MVM in the attention-head, one heater source per MRM is impractical. Therefore, this paper proposed a hybrid tuning approach by combining the tungsten-heater approach for global course tuning with the post-fabrication-trimming approach [46] for individual fine tuning. For the example shown in Fig. 1, all MRMs and RTR-PDs in the entire MVM are partitioned into three tungsten-heater regions with three corresponding heater sources, and the area of each region shall be less than 1 $mm^2$ for confining random variation within one standard deviation. Thus, when one of the heaters is heating up its own region, all transmission and absorption responses of the MRMs or RTR-PDs belonging to this region are shifted together for globally tuning the resonance wavelengths into the fine tuning-ranges of the post-fabrication trimming mechanism. The post-fabrication trimming mechanism is basically realized by implanting a section of SOI rib waveguide with Ge through a photoresist mask on the top of each MRM cavity, so each MRM resonance wavelength can be trimmed by injecting a voltage pulse to anneal this Ge rib waveguide. This annealing calibration process can be done for each MRM individually as shown in Fig. 2, and the annealing pulse width for the targeted wavelength of each MRM is reached by the iterative feedback mechanism of the MRM transmission power received by its ADC to adjust the annealing pulse generator output pulse width until converging to the targeted wavelength [46]. Although this post-fabrication trimming approach is tedious, it is thorough, reprogrammable, hardware reusable, and only consuming calibration time and power, which costs almost zero overhead during the regular MPE-MVM operations.

## V. MPE Matrix-Matrix Multiplications

To perform an MPE-MMM operation, e.g., $Y_{d\times d}\cdot Z_{d\times n} = YZ_{d\times n}$, the input d-by-n matrix $Z_{d\times n}$ with elements denoted by $z_{ij}$, i = 1 ~ d, j = 1 ~ n, can be essentially split into "n" d-by-1 column vectors as shown in Fig. 8, and each column vector individually performs MVM with matrix $Y_{d\times d}$ in an MPE-MVM unit and produce its own output column vector. After combining total "n" output column vectors from "n" parallelized MPE-MVMs, the outcome d-by-n matrix $YZ_{d\times n}$ with elements denoted by $yz_{ij}$, i = 1 ~ d, j = 1 ~ n, of the MMM operation can be obtained. Equivalently, this parallelism approach can be implemented by a time-multiplexing approach; e.g., a single MPE-MVM can process one of the input column vectors per iteration period, and electronic registers after the ADCs can collect all the output column vectors after "n" iteration cycles to form the final outcome matrix $YZ_{d\times n}$ per MMM operation. These two approaches linearly consume area and time with the column-dimension "n" of $Z_{d\times n}$, respectively. Therefore, an optimized power/area/throughput tradeoff can be done by mixing partial hardware-parallelism and partial time-multiplexing approaches based on realistic applications.

## VI. MPE Double Matrix-Matrix Multiplications

As elaborated by Eq. (3) and (6), the double matrix-matrix multiplications (D-MMM) play an important role in the attention-head computations; e.g., $X_{n\times d}\cdot Y_{d\times d}\cdot Z_{d\times n} = XYZ_{n\times n}$ with elements denoted by $xyz_{ij}$, i = 1 ~ n, j = 1 ~ n (note that each "xyz" presents a "single-word" variable, NOT "x times y times z"). Intuitively, two MPE-MMM units shown in Fig. 8 can be cascaded so that the first stage completes the operation of $Y_{d\times d}\cdot Z_{d\times n} = YZ_{d\times n}$ and then the second stage can do the $X_{n\times d}\cdot YZ_{d\times n} = XYZ_{n\times n}$. However, this consecutive MPE-MMMs architecture for the D-MMM functionality requires the intermediate multiplication result $YZ_{d\times n}$ to be converted from photonic to electronic domains and then from electronic back to photonic domains, which is a d-by-n ADC/DAC power-and-area dominant O/E/O interface with no contribution to overall computation throughput and exposing the overhead and downside of photonic computing as mentioned in Section I.

This paper proposed a single MPE-D-MMM unit to perform energy-efficient two consecutive MMM operations in the photonic domain all together, which can completely eliminate the intermediate O/E/O overhead. As shown in Fig. 9, the overall D-MMM is established by "n" double matrix-vector multiplication (D-MVM) units in parallel, and each D-MVM unit performs two consecutive MVMs at once in the photonic domain. E.g., the input vector $Z_{d\times 1,j}$ of each D-MVM unit is one of the column vectors of the input matrix $Z_{d\times n}$. After converting this input vector $Z_{d\times 1,j}$ into the photonic domain, $Z_{d\times 1,j}$ is firstly confronted by the MRM array of the matrix $Y_{d\times d}$, and then the first-stage MVM output column vector $YZ_{d\times 1,j}$ (= $Y_{d\times d}\cdot Z_{d\times 1,j}$) is generated. So far, the signal flow in Fig. 9 of the D-MVM$_j$ completing the first-stage MVM operation is exactly the same as a single MVM operation in Fig. 1. Then, each element of $YZ_{d\times 1,j}$ is duplicated by a 1-to-d O-PS for the second-stage MVM operation, i.e., $X_{n\times d}\cdot YZ_{d\times 1,j}$. Note that each element of $YZ_{d\times 1,j}$ already contains all wavelengths ($\lambda_j$, j = 1 ~ d) carrying their own weight factors. Therefore, to further perform MVM with $YZ_{d\times 1,j}$, the control-voltage of each element ($x_{ij}$, i = 1 ~ n, j = 1 ~ d) of $X_{n\times d}$ requires a RTR-based modulator (RTM) to influence the amplitudes of all wavelengths in each element of $YZ_{d\times 1,j}$ all together. This RTM is basically exactly the same as the RTR-PD discussed in Section IV.E, including the footprint, FSR, WDM spacing, and resonance wavelengths, but each RTM is driven by $x_{ij}$ for broadband light-wave power modulations instead of detections. After the broadband modulations for element-to-element products of $X_{n\times d}$ and $YZ_{d\times 1,j}$, the following "n" d-to-1 O-MUXs are required to perform the summations of element-to-element products. Finally, the outputs of these "n" d-to-1 O-MUXs represent the outcome vector of the D-MVM operation, $XYZ_{n\times 1,j}$ (= $X_{n\times d}\cdot YZ_{d\times 1,j}$). With the parallel architecture shown in Fig. 9, the "n" D-MVM units can accomplish the whole D-MMM operation ($X_{n\times d}\cdot Y_{d\times d}\cdot Z_{d\times n}$ = $XYZ_{n\times n}$) fully in the photonic domain for eliminating the intermediate O/E/O overhead and maintaining negligible computation latency and power consumption with the downside of area consumption and calibration effort due to additional "n×d" RTMs, "n×d" R2R-DACs, "d" O-PSs and "d" O-MUXs per D-MVM unit.

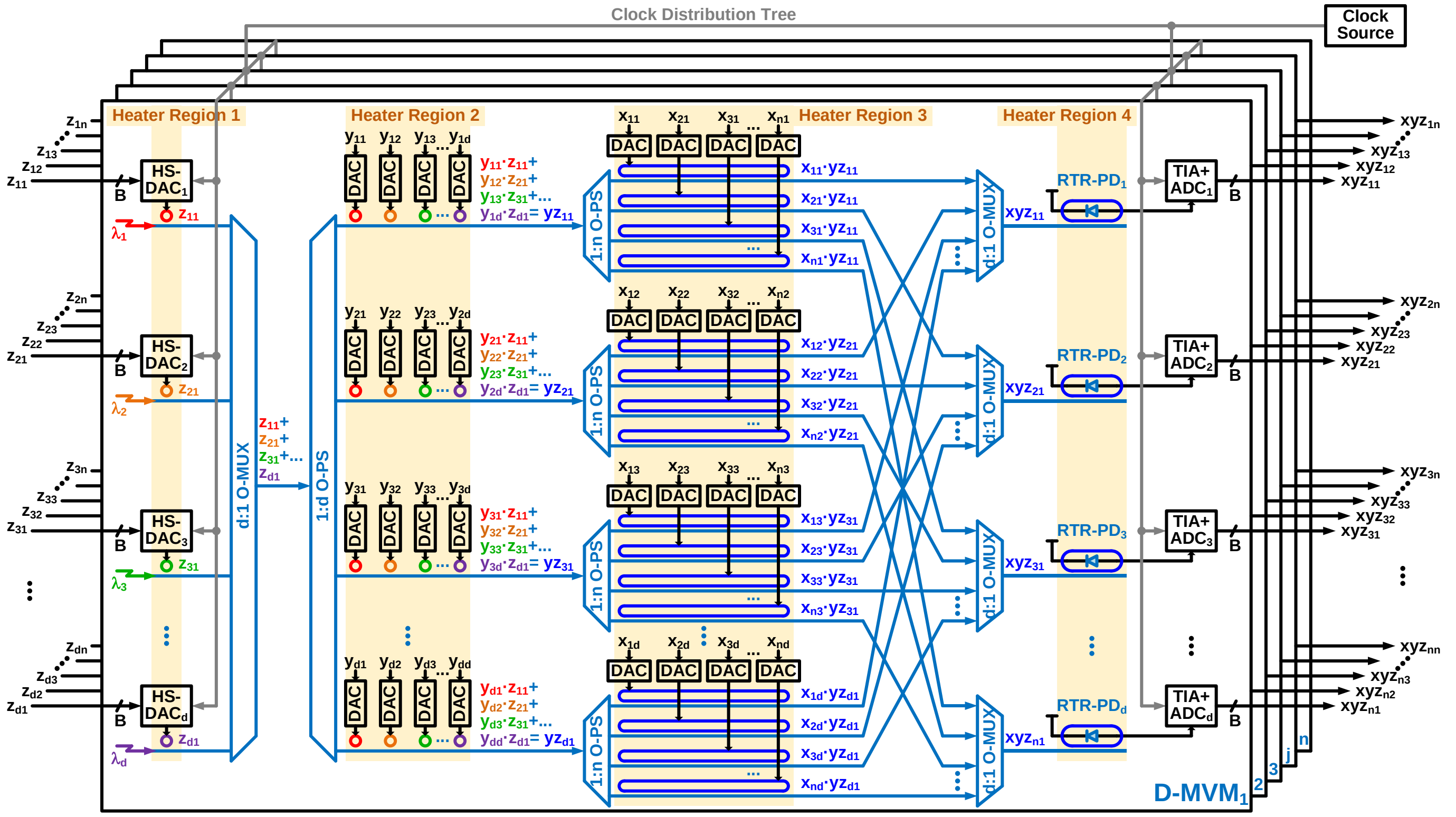

Fig. 9. The system block diagram of the MPE-D-MMM linear-algebra accelerator implemented by the D-MVM parallelism approach. Note that each "yz" presents a "single-word" variable, NOT "y times z"; each "xyz" also presents a "single-word" variable, NOT "x times y times z".

## VII. Performance Summary and Conclusion

The primary performance metrices of high-performance computing systems include computation throughput (TMAC/s), computation density (TMAC/s/mm$^2$), and energy consumption (fJ/MAC) [15], [18], [23]. By consolidating the power, area, and operating frequency of the proposed building blocks elaborated and analyzed in Section IV, the detailed power/area breakdowns and computing performance metrics of the MPE-MVM accelerator with its entire single-chip hardware and standalone MVM functionality are summarized in Table II to practically cover the overhead of the MPE integrations, conversions, and calibrations, including all electronic/photonic SoC building blocks with complete DACs/ADCs, on-chip digital interface, clock distribution, on-chip calibration hardware, laser injection power, and heater power, instead of only considering the photonic power/area in the performance evaluations [15], [18], [23]. For the same reason of thoroughness in practical hardware realization, the performance metrics of Google Tensor Processing Units (TPU) [47], [48] are listed in Table II as well for the comparison purpose since these TPUs are also complete and fully integrated SoCs but implemented by digital application-specific integrated-circuits

TABLE II
PERFORMANCE ESTIMATIONS OF MPE-MVM VS. ASIC-MVM

| **MPE-MVM** in 45-nm Monolithic SiPh [This Work] | Vector Dimen. "d" | Laser Power (mW) | Heater Power (mW) | SoC Power* (mW) | SoC Area ($mm^2$) | Data Precision (bit) | Clock Rate (GHz) | Computation Throughput TOPS/s | Computation Throughput TMAC/s | Computation Density (TMAC/s/$mm^2$) | Energy Consumption (fJ/MAC)* |
|---|---|---|---|---|---|---|---|---|---|---|---|
| | 8 | 31.6 | 40.8 | 99.6 | 0.10 | 4 | 2 | 0.256 | 0.128 | 1.26 | 777.8 |
| | 16 | 64.3 | 79.2 | 198.7 | 0.33 | 4 | 2 | 1.024 | 0.512 | 1.56 | 388.0 |
| | 32 | 130.7 | 156.0 | 400.7 | 1.14 | 4 | 2 | 4.096 | 2.048 | 1.80 | 195.6 |
| | 64 | 265.6 | 309.6 | 818.0 | 4.16 | 4 | 2 | 16.384 | 8.192 | 1.97 | 99.8 |
| | 128 | 539.9 | 616.8 | 1701.1 | 15.77 | 4 | 2 | 65.536 | 32.768 | 2.08 | 51.9 |
| | 256 | 1097.3 | 1231.2 | 3653.3 | 61.12 | 4 | 2 | 262.144 | 131.072 | 2.14 | 27.9 |

| **ASIC-MVM** Google TPUv1 in 28-nm CMOS [47] | Vector Dimen. "d" | SoC Idle Power (mW) | SoC Busy Power (mW)† | SoC Area ($mm^2$) | Data Precision (bit) | Clock Rate (GHz) | Computation Throughput TOPS/s | Computation Throughput TMAC/s | Computation Density (TMAC/s/$mm^2$) | Energy Consumption (fJ/MAC)† |
|---|---|---|---|---|---|---|---|---|---|---|
| | 256 | 28000 | 40000 | 331 | 8 | 0.7 | 91.76 | 45.88 | 0.14 | 871.8 |

| **ASIC-MVM** Google TPUv4 in 7-nm CMOS [48] | Vector Dimen. "d" | SoC Idle Power (mW) | SoC Busy Power (mW)† | SoC Area ($mm^2$) | Data Precision (bit) | Clock Rate (GHz) | Computation Throughput TOPS/s | Computation Throughput TMAC/s | Computation Density (TMAC/s/$mm^2$) | Energy Consumption (fJ/MAC)† |
|---|---|---|---|---|---|---|---|---|---|---|
| | 256 | 55000 | 78571‡ | 400 | 8 | 1.05 | 137.62 | 68.81 | 0.17 | 1141.9‡ |

* Including all photonic/electronic devices/circuits power consumption, heater power, and laser injection power on the single monolithic SiPh chip.
† Including all electronic digital circuits power consumption on the single CMOS chip in the Busy-mode.
‡ The Busy-mode power of TPUv4 is estimated by its Idle-mode power (= 55 W) and Busy-vs.-Idle power ratio of TPUv1 (= 1.43).

(ASIC). Note that the performance metrics of the MPE-MMM and MPE-D-MMM accelerators can be reasonably estimated according to those of the MPE-MVM accelerator because of the dimension scalability and parallelism architecture described in Section V and VI.

The performance scalability with the input vector dimension "d" of the MPE-MVM accelerator listed in Table II shows that the power/area overhead of MPE-MVM is getting leveraged by the negligible photonic computing latency and power consumption when "d" is scaling up. In the cases of "d" ≥ 8, the MPE-MVM accelerator outperforms the ASIC counterparts in both computation-density and energy-consumption. In particular with the future advances in scaling the optical comb generations to "d" = 256, the MPE-MVM accelerator can exhibit about 12.6× computation-density and 40.9× energy-efficiency superiority over the advanced ASIC-MVM accelerator (TPUv4) with the downside of lower data precision due to the overhead of "analog" computing in photonics. Finally, it is important to note that many of novel SiPh devices and circuits are still being discovered and engineered for future foundry manufacturing; thus, the performance metrics of the MPE accelerators can be further improved with the development of next-generation SiPh process technology.

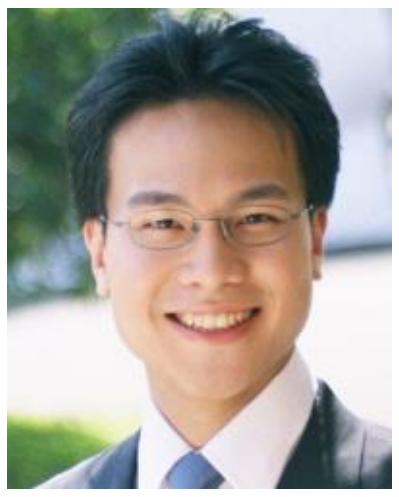

**Tzu-Chien Hsueh** (Senior Member, IEEE) received the Ph.D. degree in electrical and computer engineering from the University of California, Los Angeles, CA, in 2010. From 2001 to 2006, he was a Mixed-Signal Circuit Design Engineer in Hsinchu, Taiwan. From 2010 to 2018, he was a Research Scientist in Intel Lab Signaling Research and an Analog Engineer in Intel I/O Circuit Technology, Hillsboro, Oregon. Since 2018, he has been an Assistant Professor in electrical and computer engineering at the University of California, San Diego (UCSD). His research interests include wireline electrical/optical transceivers, clock-and-data recovery, data-conversion circuits, on-chip performance measurements/analyzers, and digital/mixed signal processing techniques.

Dr. Hsueh was a recipient of multiple Intel Division and Academy Awards from 2012 to 2018, the 2015 IEEE Journal of Solid-State Circuits (JSSC) Best Paper Award, the 2020 NSF CAREER Award, and the 2022 UCSD Best Teacher Award. He served on the Patent Committee for Intel Intellectual Property (Intel IP) and the Technical Committee for Intel Design & Test Technology Conference (DTTC) from 2016 to 2018. Since 2018, he has served on the Technical Program Committee for IEEE Custom Integrated Circuits Conference (CICC) and the Guest Associate Editor for IEEE Solid-State Circuits Letters (SSC-L).

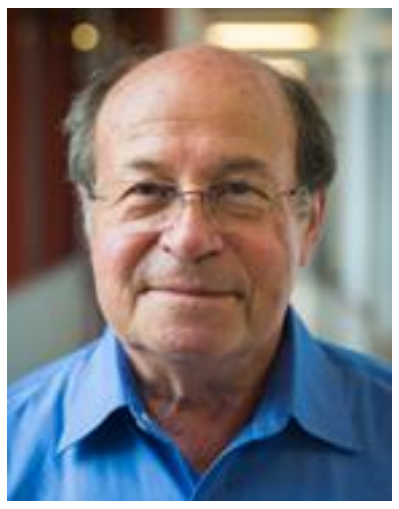

**Yeshaiahu (Shaya) Fainman** (Fellow, IEEE) is an inaugural ASML/Cymer Chair Professor of Advanced Optical Technologies and Distinguished Professor in Electrical and Computer Engineering (ECE) at the University of California, San Diego (UCSD). He received M. Sc and Ph. D degrees from Technion-Israel Institute of Technology in 1979 and 1983, respectively. He is directing research of the Ultrafast and Nanoscale Optics group at UCSD and made significant contributions to near field optical phenomena, nanoscale science and engineering of ultra-small, sub-micrometer semiconductor light emitters and nanolasers, inhomogeneous and meta-materials, nanophotonics and Si Photonics. His current research interests are in near field optical science and technology with Si Photonics applications to information technologies and biomedical sensing. He contributed over 340 manuscripts in peer review journals and over 560 conference presentations and conference proceedings. He contributed to editorial and conference committee works of various scientific societies including IEEE, SPIE and OPTICA. He is a Fellow of OPTICA (former OSA), Fellow of the IEEE, Fellow of the SPIE, and a recipient of the Miriam and Aharon Gutvirt Prize, Lady Davis Fellowship, Brown Award, Gabor Award, Emmett N. Leith Medal, Joseph Fraunhofer Award/Robert M. Burley Prize and OPTICA Holonyak Award.

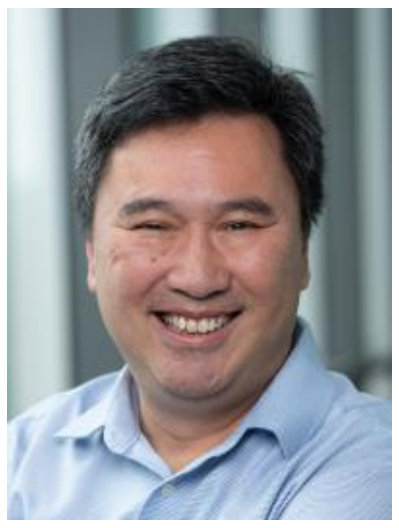

**Bill Lin** (Senior Member, IEEE) received the B.S., M.S., and Ph.D. degrees in electrical engineering and computer sciences from the University of California at Berkeley, Berkeley, CA, USA, in 1985, 1988, and 1991, respectively. He is currently a Professor in electrical and computer engineering with the University of California, San Diego, La Jolla, CA, USA, where he is actively involved with the Center for Wireless Communications (CWC), the Center for Networked Systems (CNS), and the Qualcomm Institute in industry-sponsored research efforts. Dr. Lin's research has led to more than 200 journal and conference publications, including a number of Best Paper awards and nominations. He also holds five awarded patents. He was the General Chair and on the executive and technical program committee of many IEEE and ACM conferences, and he was Associate and Guest Editors for several IEEE and ACM journals.